\documentclass[twocolumn,aps,pra,groupedaddress,amsmath,amssymb,showpacs]{revtex4-1}
\usepackage{graphicx,amsmath,amssymb,epsfig,color}
\usepackage{bm}
\usepackage{float,footnote,comment,ulem}
\usepackage{txfonts}
\usepackage[colorlinks,	linkcolor=blue,	anchorcolor=blue,citecolor=blue,bookmarksnumbered]{hyperref}

\begin{document}

\title{Facilitation Induced Transparency and Single-Photon Switch with Dual-Channel Rydberg Interactions}
\author{Yao Ding$^{1}$}
\author{Zhengyang Bai$^{1}$}
\email{zhybai@lps.ecnu.edu.cn}
\author{Guoxiang Huang$^{1}$}
\author{Weibin Li$^{2}$}
\affiliation{$^1$State Key Laboratory of Precision Spectroscopy, East China Normal University, Shanghai 200062, China\\
             $^2$School of Physics and Astronomy, and Centre for the Mathematics and Theoretical Physics of Quantum Non-equilibrium Systems, University of Nottingham, Nottingham, NG7 2RD, UK
             }
\date{\today}

\begin{abstract}
We investigate facilitation induced transparency (FIT) enabled by strong and long-range Rydberg atom interactions between two spatially separated optical channels. In this setting, the resonant two-photon excitation of Rydberg states in a target channel is conditioned by a single Rydberg excitation in a control channel. Through the contactless coupling enabled by the Rydberg interaction, the optical transparency of the target channel can be actively manipulated by steering the optical detuning in the control channel. By adopting a dressed-state picture, we identify two different interference pathways, in which one corresponds to Rydberg blockade and an emergent one results from facilitation. We show that the FIT is originated from the Rydberg interaction and the quantum interference effect between the two pathways, which is different from conventional electromagnetically induced transparency realized by single-body laser-atom coupling.  We find that the FIT in such a dual-channel setting is rather robust, insensitive to changes of systemic parameters, and can be generalized to multi-channel settings. Moreover, we demonstrate that such a FIT permits to realize controllable single-photon switches, which also paves a route to detect Rydberg facilitation by using optical absorption spectra. Our study contributes to current efforts in probing correlated many-body dynamics and developing single-photon quantum devices based on Rydberg atom ensembles.
\end{abstract}

\maketitle
\section{Introduction}


Cold gases of Rydberg atoms have emerged as a versatile platform for studying quantum nonlinear optics, non-equilibrium  statistical physics, and quantum simulation of strongly interacting many-body systems~\cite{saffman_quantum_2010,MURRAY_quantum_2016,Adams_Rydberg_2019,
browaeys_many-body_2020}, including antiferromagnetic phase transition~\cite{Lee_Antiferromagnetic_2011, bernien2017probing}, quantum many-body scars~\cite{Interacting_Igor_2012,turner2018weak,bluvstein_controlling_2021}, and Heisenberg XYZ spin model~\cite{Floquet_Geier_2021}. At the same time, Rydberg atoms are of technological importance, which allow to create collectively encoded qubit~\cite{Lukin_Dipole_2001,Spong_Collectively_2021},  entanglement~\cite{Preparation_Carr_2013, Jau_Entangling_2016}, photonic or neutral-atom gates for quantum information processing~\cite{saffman_quantum_2010,Khazali_Photon_2015,Infidelity_Liu_2021,
Hybrid_Li_2022}, precision measurements~\cite{sedlacek_microwave_2012,Jing_microwave_2020}, and so on. The enabled fundamental research and practical applications are rooted largely by the fact that Rydberg atoms offer strong and long-ranged interactions. In addition, their internal and external states can be actively manipulated under current experimental conditions~\cite{browaeys_many-body_2020,Vasilyev_Monitoring_2020}.

Rydberg many-body physics is largely based on two mechanisms: Rydberg blockade and facilitation (antiblockade), which can be realized by tuning the excitation laser on resonance with  atomic transitions, and for matching the Rydberg interaction induced energy shifts~\cite{su_rydberg_2020}, respectively. The former prevents multiple Rydberg excitations in the vicinity of a Rydberg atom, which benefits to the creation of quantum correlation and entanglement between atoms~\cite{Preparation_Carr_2013, EIT_Li_2014}. The facilitation, opposite to the blockade, allows the formation of Rydberg clusters~\cite{Ates_Antiblockade_2007, Malossi_Full_2014, Statistics_Schempp_2014, Khazali_Photon_2015,  Marcuzzi_Facilitation_2017,Localization_Liu_2022}; it has been a central building block of non-equilibrium phase transitions, such as aggregation~\cite{Statistics_Schempp_2014, Out_Lesanovsky_2014}, epidemic spreading~\cite{Epidemic_Espigares_2017,Wintermantel_Epidemic_2021}, and self-organized criticality~\cite{Signatures_Helmrich_2020, Ding_Phase_2020}, etc.. Experimentally, Rydberg facilitation can be detected via absorption imaging~\cite{Signatures_Helmrich_2020} or direct ionization~\cite{amthor_evidence_2010,Gutierrez_Experimental_2017,
Bai_antiblockade_2020}. However, although these methods have high efficiency, they typically damage many-body coherence after experimental operations~\cite{Coherent_Mohapatra_2007}.

In this work, we theoretically investigate the optical property of a dual-channel Rydberg ensemble~\cite{Two_He_2014,Daniel_Electromagnetically_2015,Busche_Contactless_2017,thompson_symmetry-protected_2017,
Khazali_Polariton_2019,Two_Chen_2021}, and illustrate that the facilitation mechanism can lead to an effective interaction to probe photons, which can be directly measured by using atomic absorption spectrum. The system we consider consists of two optical channels that are remotely separated in space, in which target channel $A$ is resonantly excited to a Rydberg state via electromagnetically induced transparency (EIT)~\cite{pritchard_cooperative_2010} and affected by control channel $B$ via long-ranged Rydberg interactions. Such a setting is available to realize with present-day experimental techniques~\cite{Busche_Contactless_2017,thompson_symmetry-protected_2017,Two_Chen_2021}.

We show that a transparency window for probe photons opens through adjusting the detuning of the channel $B$, and this transparency window is sensitive to the Rydberg interaction. By using numerical and dressed-state calculations, we reveal many-body quantum pathways and characterize an interaction-dependent scaling function. A key finding is that the two resonant transition pathways are respectively originated from the Rydberg blockade and facilitation; it is the quantum interference effect between the two pathways that leads to the facilitation induced transparency (FIT) of the probe photons. The relevant model can moreover be generalized to Rydberg atomic gases with multiple optical channels.

The results on the FIT described above may have potential applications in all-optical quantum information processing. To realize optical quantum devices at single photon level, up to now various quantum systems have been suggested to implement  optical switches or transistors, such as single molecules~\cite{single_Hwang_2009}, quantum dots~\cite{Low_Bose_2012, Ultrafast_Englund_2012, Optical_Loo_2012, Ultrafast_Volz_2012} and neutral atoms~\cite{Efficient_Bajcsy_2009,Shea_Fiber_2013}.
However, due to the saturable absorption of single quantum emitters, it is difficult to realize a sizable optical non-linearity at single-photon level. To enhance light-matter interactions, one way is to use high-finesse optical resonators to confine photons to small mode volumes~\cite{Chen_All_2013, Quantum_Chang_2014}.
At variance with single-body quantum systems, strong and long-range interactions between Rydberg atoms can be effectively  mapped onto photon-photon interactions via two-photon scheme. This provides an alternative avenue to design single-photon switching in {\it free space}~\cite{Single_Baur_2014, Single_Gorniaczyk_2014, tiarks_single-photon_2014}.
We demonstrate that the system proposed here can be used to design contactless single-photon switches, which is different from early studies, where Rydberg single-photon switches were realized in the blockade regime~\cite{Single_Baur_2014, Single_Gorniaczyk_2014, tiarks_single-photon_2014, Weibin_Coherence_photon_switch2015, Many_Murray_2016,Experimental_Yu_2020,Two_Chen_2021}. By employing the presented dual-channel system, one can easily launch both facilitated and blockade single-photon switch by manipulating the control channel. 

Moreover, different from that of stationary Rydberg single-photon switches reported before, the efficiency of the FIT-based photon switch suggested here can be controlled dynamically. The use of the dual-channel setting brings more potential benefits. Here Rydberg atoms in the control and target channels are spatially separated. For example, complicated atomic and molecular processes could be avoided, which play roles in the Rydberg blockade in atomic ensembles~\cite{derevianko_effects_2015}. As we will show later, Rydberg populations in the FIT is relatively weak. This could mitigate the formation of ultralong-range Rydberg  molecules~\cite{bendkowsky_observation_2009,li_homonuclear_2011} and Rydberg polarons~\cite{camargo_creation_2018} due to collisions between ground state and Rydberg atoms~\cite{Bai_Self_2020}. These features mean that coherence of the atom could be increasingly protected in the FIT regime.

The remainder of the article is arranged as follows. In Sec.~\ref{sec_model}, we introduce the Rydberg dual-channel model based on a master equation description. The level scheme and typically parameters are given. In Sec.~\ref{sec_FIT}, we reveal the FIT mechanism through analyzing a two-atom model both analytically and numerically. In Sec.~\ref{coherence}, coherence of the probe field is obtained numerically. By comparing with the approximate result, we reveal the importance of two-channel correlations in the emergence of FIT.  In Sec.~\ref{sec_dressed}, we present an analytical dressed-state approach on the model and show the scaling of FIT as a function of the inter-channel interaction. In Sec.~\ref{sec_switch}, based on the FIT we discuss controllable single-photon switch protocol. Finally,  Sec.~\ref{conclusion} gives a summary of the main results obtained in this work.

\section{Dual-channel Rydberg model}\label{sec_model}

\begin{figure}[htb]
\centering
\includegraphics[width=0.47\textwidth]{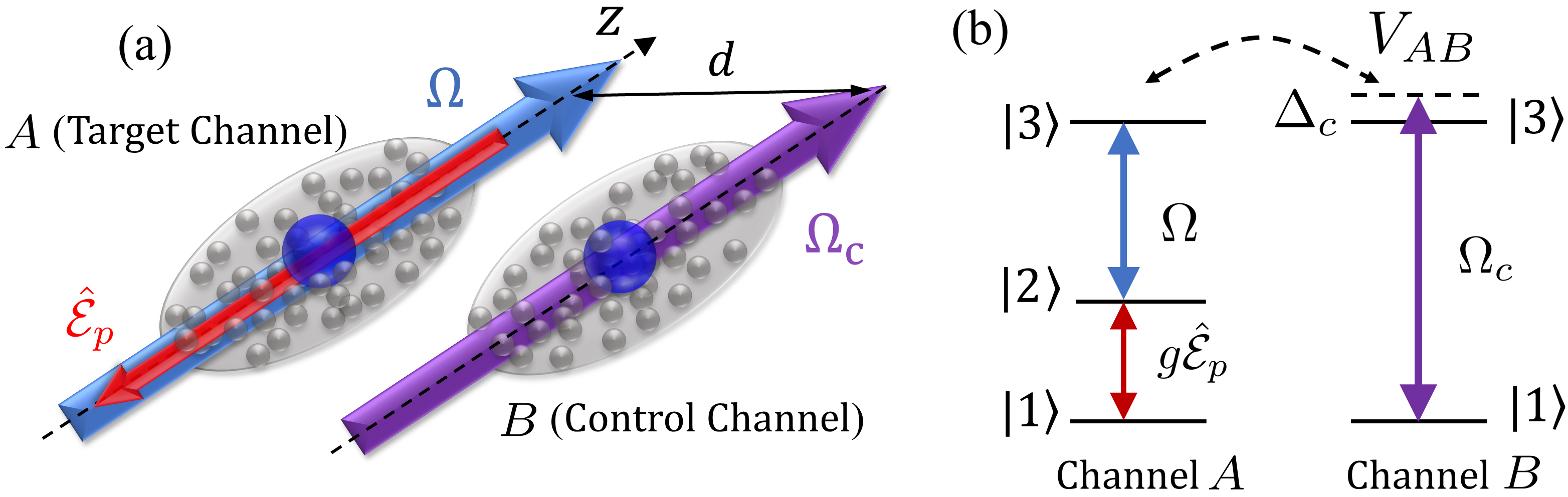}\\
\caption{\footnotesize\textbf{Scheme of dual-channel Rydberg atomic gas for realizing FIT}. Two atomic ensembles, i.e. target channel $A$ and control channel $B$, are prepared in parallel and separated by distance $d$.
(a)~A weak probe field $\hat{{\cal E}}_p$ propagates in the channel $A$, whose dynamics is facilitated by the Rydberg excitation in the channel $B$. Long-range inter-channel Rydberg interaction permits a contactless control for the two channels and hence the propagation of the probe photon.
(b)~Level diagram and excitation scheme. Atoms in the channel $A$  have three quantum states (i.e. $|1\rangle$, $|2\rangle$, $|3\rangle$; $|3\rangle$ is a Rydberg state) forming a Rydberg-EIT, where resonant transitions $|1\rangle\leftrightarrow|2\rangle$ and $|2\rangle\leftrightarrow|3\rangle$ are respectively provided by weak probe field $\hat{\cal E}_p$ and strong coupling field (Rabi frequency $\Omega$). In the channel $B$,  a control field with Rabi frequency $\Omega_c$ drives the transition between the ground state $|1\rangle$ to the Rydberg state $|3\rangle$ with detuning $\Delta_c$. $V_{AB}$ is the van der Waals interaction between the atoms in the two channels.
} \label{Model}
\end{figure}

The system we consider consists of two ensembles of cold atoms driven by channel-dependent laser fields. The two ensembles are prepared in parallel, elongated traps along $z$ direction, and separated by distance $d$; see Fig.~\ref{Model}(a).
They form two optical channels, i.e. a target channel $A$ (containing $N_A$ atoms) and a control channel $B$ ($N_B$ atoms).
The atomic level structure and the optical excitation scheme are illustrated schematically in Fig.~\ref{Model}(b). We assume that, in the target channel $A$, a weak probe laser field $\hat{\cal E}_p(z)$ [Rabi frequency $\hat{\Omega}_p(z)=g\hat{\cal E}_p (z)$] resonantly couples the ground state $|1\rangle$ and the excited state $|2\rangle$, and
a coupling laser field (Rabi frequency $\Omega$) resonantly couples  the state $|2\rangle$ and the Rydberg state $|3\rangle$, which forms a scheme of conventional Rydberg-EIT~\cite{Coherent_Mohapatra_2007}. Here $g=\mu_{21}\sqrt{\omega_p/(2\hbar\epsilon_0V_p)}$ is the atom-field coupling constant, with $\omega_p$ the central angular frequency of the probe field, $V_p$ the quantization
volume of the probe field, and $\mu_{21}$  the electric-dipole matrix element associated with the transition between $|1\rangle$ and $|2\rangle$.
In the control channel $B$, atomic levels $|1\rangle$ and $|3\rangle$ are coupled off-resonantly by another control laser field (Rabi frequency $\Omega_c$) with detuning $\Delta_c$, which plays a critical role for realizing the FIT. We assume that both $\Omega$ and $\Omega_c$ are strong enough, thus they can be taken as classical fields and their depletion are negligible.

Under the rotating wave approximation, by setting $\hbar=1$
the Hamiltonian for the channels $A$ and $B$ are respectively given by
\begin{subequations}
\begin{eqnarray}
&& \hat{H}_{A} = -\frac{1}{2}\sum_{j\in A}  \left[\hat{\Omega}_p\hat{\sigma}_{21}(z_j)+\Omega\hat{\sigma}_{32}(z_j)+{\rm H.c.}\right],\\
&& \hat{H}_{B}=-\sum_{j\in B}\left[\frac{\Omega_c}{2}\hat{\sigma}_{31}(z_j)-\Delta_c\hat{\sigma}_{33}(z_j)+{\rm H.c.}\right],
\end{eqnarray}
\end{subequations}
with $\hat{\sigma}_{ab}(z_j)\equiv|a_j\rangle\langle b_{j}|$\,  ($z_j$ is the position of $j$th atom in the respective ensemble) and H.c. representing Hermitian conjugate of the preceding terms.
We assume that the spatial extension of each atomic ensemble is small, so that multiple Rydberg excitations within each ensemble is prohibited by the strong and long-ranged interaction between Rydberg states. However, there exists an inter-channel Rydberg interaction between the target channel $A$ and the control channel $B$ (with the distance $d$ typically Rydberg blockade radius $\sim 5-10$\,$\mu$m), which leads to the following inter-channel interaction Hamiltonian
\begin{equation}
H_{\rm int}=\sum_{j\in A, l\in B}V(z_{jl})\hat{\sigma}_{33}(z_j)\hat{\sigma}_{33}(z_l),
\end{equation}
Here we have assumed that the target and control channels
are both excited to $|nS\rangle$ states (or $|nD\rangle$ states). Two Rydberg channels interact via the van der Waals (vdW) interaction  $V(z_{jl})=-C_6/[(z_j-z_l)^2+d^2]^3$, with $C_6$ the  dispersion coefficient.
Note that, except for the vdW interaction, a resonant dipole-dipole interaction can also be engineered by coupling the ensembles to Rydberg states $|S\rangle$ and $|P\rangle$, which may lead to spin-exchange between remote Rydberg channels~\cite{EIT_Li_2014,Daniel_Electromagnetically_2015,thompson_symmetry-protected_2017,Khazali_Polariton_2019}. This is a topic deserving to be explored, but here we focus our consideration only on the case of the vdW interaction.



The dynamics of the system including the two channels is governed by the master equation,
\begin{eqnarray}\label{Master}
\frac{d\rho}{dt}=-i\left[H,\rho\right]+D(\rho)+D_d(\rho),
\end{eqnarray}
where $\rho(t)$ is density matrix, and $\hat{H}=\hat{H}_{A}+\hat{H}_{B}+H_{\rm int}$ is the total Hamiltonian of the system. In Eq.~(\ref{Master})
\begin{eqnarray}
D(\rho)=&&\sum_{\mu=A, B}\sum_{j\in\mu}\Gamma_{ba}^\mu\left\{\hat{\sigma}_{ba}(z_j)\rho\hat{\sigma}_{ab}(z_j)
\right.\nonumber\\
&&\left. -\frac{1}{2}\left[\rho\hat{\sigma}_{ab}(z_j)\hat{\sigma}_{ba}(z_j)
+\hat{\sigma}_{ab}(z_j)\hat{\sigma}_{ba}(z_j)\rho\right]\right\},
\end{eqnarray}
describes the respective dissipation processes from $|a\rangle \to |b\rangle$ with the rates $\Gamma_{ba}^\mu=\Gamma_{12}^A$, $\Gamma_{23}^A$, and $\Gamma_{13}^B$;
\begin{equation}
D_d(\rho)=\sum_{\mu=A, B}\sum_{j\in\mu}\gamma_{a}^\mu\left\{2\hat{\sigma}_{aa}(z_j)\rho\hat{\sigma}_{aa}(z_j)
-\left[\rho\hat{\sigma}_{aa}(z_j)+\hat{\sigma}_{aa}(z_j)\rho\right]\right\},
\end{equation}
gives the dephasing of the atomic coherence (originated from atomic collisions, residue Doppler effect, dipole-dipole interaction between the Rydberg atoms, finite laser linewidth, etc.), with rates $\gamma_{a}^\mu$ ($\mu=A, B$; $a=2, 3$).
In the following, for the convenience of numerical calculations, we measure all atomic parameters (e.g. $\Omega_c$, $\Omega_p$, $\Delta_c$) in unit of $\Gamma_{12}^A$ (typically, $\Gamma_{12}^A\simeq2\pi\times6.06$\,MHz for $|5P_{3/2}\rangle$ state in Rb atom; $\Gamma_{12}^A\simeq2\pi\times5.22$\,MHz for $|6P_{3/2}\rangle$ state in Cs atom~\cite{steck_alkali_nodate}).

In this setting, the inter-channel interaction is the central element in the Hamiltonian, with which the dynamics of the target channel $A$ can be actively manipulated by varying the relevant atomic parameters in the control channel $B$ (i.e., $\Omega_c$, $\Delta_c$).  Note that
the control channel $B$ is adopted with a two-photon Raman scheme. Due to the large single-photon detuning, the channel $B$ can be described by an effective two-level system. Though this is a simplified approach, it can demonstrate the essential physics inherent in the model. Dynamics of the two-photon scheme in channel $B$ is presented in Appendix~\ref{ap1} and will be briefly discussed below in Fig.~\ref{Scaling}(c).

%

\section{Facilitation Induced Transparency}~\label{sec_FIT}

In order to demonstrate the FIT and to reveal its basic physical mechanism, we first consider a simple two-atom scenario [i.e. one atom (atom $A$) is in the channel $A$ and another one (the atom $B$) is in the channel $B$]
that represents the minimal model of the dual-channel setting~\cite{EIT_Li_2014,Daniel_Electromagnetically_2015}.

With such a consideration, the effective Hamiltonian is reduced into the form of $\hat{H}=\hat{H}_A+\hat{H}_B+\hat{H}_{\rm int}$, where $\hat{H}_A=-(\hat{\Omega}_{p}\hat{\sigma}_{21}^A/2+\Omega\hat{\sigma}_{32}^A/2)+{\rm H.c.}$, $\hat{H}_B=-\Omega_{c}\hat{\sigma}_{31}^B/2+\Delta_c\hat{\sigma}_{33}^B+{\rm H.c.}$, and $H_{\rm int}=V_{AB}\hat{\sigma}_{33}^{A}\hat{\sigma}_{33}^{B}$.
Here $\hat{\sigma}_{ab}^{A}=\hat{\sigma}_{ab}(z_A)$, $\hat{\sigma}_{ab}^{B}=\hat{\sigma}_{ab}(z_B)$,
the two-body interaction is given by $V_{AB}=-C_6/d^6$.
Note that under the EIT condition the probe field has negligible attenuation and can be approximately treated as a classical one~\cite{Gorshkov_Photon_2007}, hence for simplicity we replace  $\hat{\Omega}_p$ by $\Omega_p$ in the approach of this section. The spatial-temporal dynamics of the quantized probe field at a level of single photon~\cite{fleischhauer_quantum_2002,Weibin_Coherence_photon_switch2015,
Quantum_Zhang_2021} in high density atom gases will be analyzed in section~\ref{sec_switch}.

\subsection{Optical coherence and inter-channel correlation}~\label{coherence}

The response character of the system to the probe field is characterized by the atomic coherence between the states $|1\rangle$ and $|2\rangle$ in the channel $A$, i.e. $\rho_{21}^A$, where $\rho_{ab}^\mu\equiv {\rm Tr}(\rho \sigma_{ba}^\mu)$ ($\mu=A,B$) are one-body reduced density matrix (DM) elements. The absorption (refraction) of the probe field is determined by the imaginary (real) part Im($\rho_{21}^A$)\, [Re($\rho_{21}^A$)\,] of $\rho_{21}^A$. The corresponding optical susceptibility will be given in section~\ref{sec_switch}; see Eq.~(\ref{susceptibility}) below. In the following calculations, we shall examine in particular the behavior of $\rho_{21}^A$ when adjusting the detuning $\Delta_c$ in the channel $B$, which is one of the system parameters easy to control experimentally.

\begin{figure}[htb]
\centering
\includegraphics[width=0.5\textwidth]{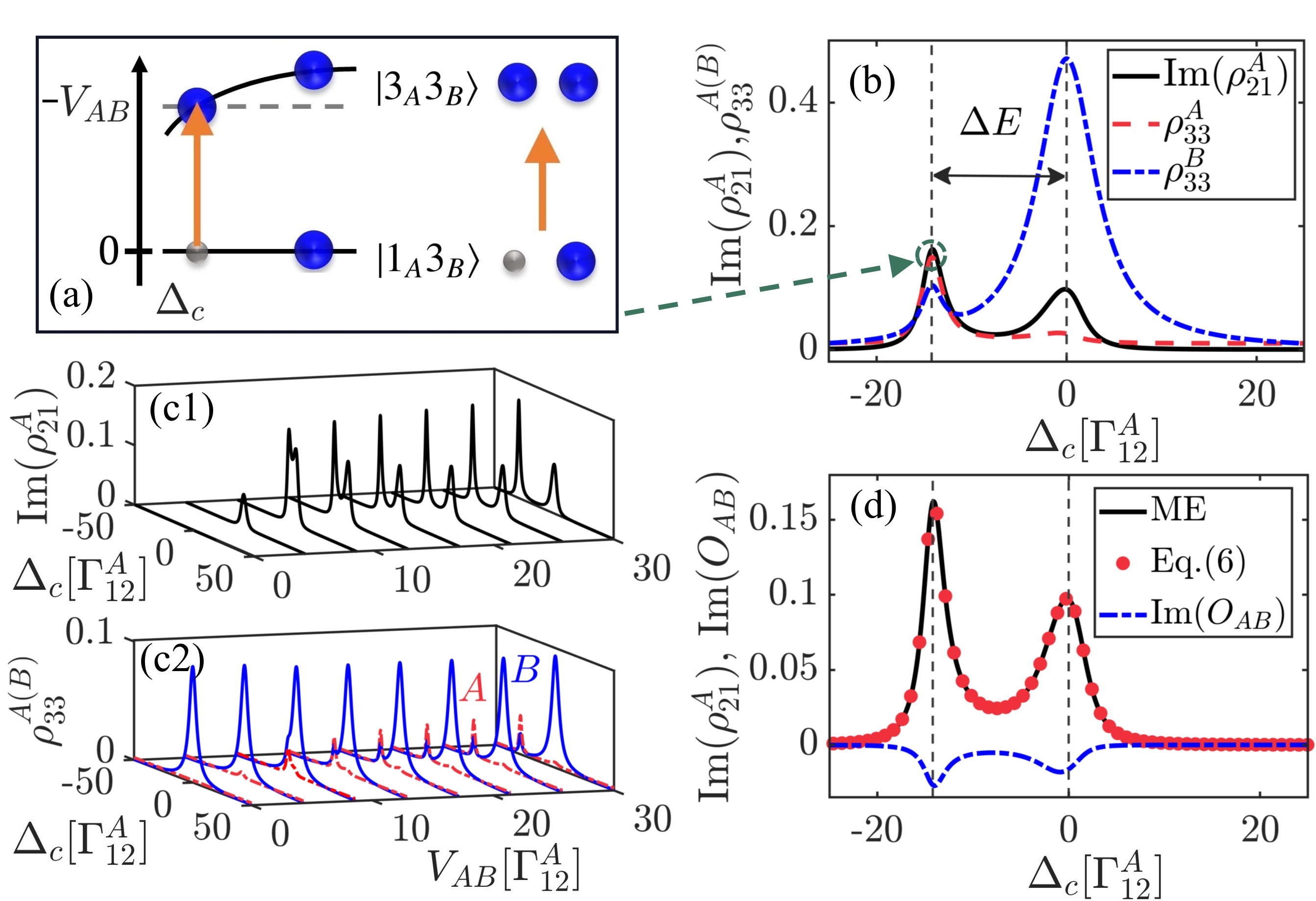}\\
\caption{\footnotesize\textbf{Optical coherence and inter-channel correlation in the two-atom system.}
(a) Level diagram and excitation scheme between the Rydberg blockade state $|1_A 3_B\rangle$ (occurring at $\Delta_c=0$) to the facilitated state $|3_A3_B\rangle$ (occurring at $\Delta_c=-V_{AB}$) of the two-atom model. For more details, see the text.
(b) FIT transparency window induced by the inter-channel interaction $V_{AB}$, as seen in the absorption spectrum Im($\rho_{21}^A$) as a function of $\Delta_c$. $\Delta E$ is the energy separation between centers of the two peaks of the FIT spectrum.
(c1) Im($\rho_{21}^A$) as a function of $\Delta_c$ and $V_{AB}$.  Im($\rho_{21}^A$) exhibits a single peak for smaller $V_{AB}$; it however splits into two peaks for larger $V_{AB}$. The dip between the two peaks becomes wider and deeper as $V_{AB}$ increases, similar to the behavior of EIT in three-level systems. Corresponding Rydberg excitations are shown in (b)
(dashed red line for $\rho_{33}^{A}$; dot-dashed blue line for $\rho_{33}^{B}$), and also in (c2) which shows that the channel $A$ is fully blockaded independent of $V_{AB}$ (solid blue line) for $\Delta_c=0$ but $\rho_{33}^A$ is facilitated around $\Delta_c=-V_{AB}$ (dashed red line).
(d) Analytical result (dotted red line) given by Eq.~(\ref{Correlation}), which captures well with the one by solving the master equation (ME) numerically ( solid black line). The quantum correlation $O_{AB}$ (dot-dashed blue line) is significant for the FIT effect. The parameters used are $\Omega=\Omega_c=5\Gamma_{12}^A$, $\Omega_p=0.5\Gamma_{12}^A$, $\Gamma_{23}^A=10^{-3}\Gamma_{12}^A$, and $\Gamma_{13}^B=\Gamma_{12}^A$; in panels (b) and (d), $V_{AB}=15\Gamma_{12}^A$.
} \label{FIT_Windows}
\end{figure}
Fig.~\ref{FIT_Windows}(b) shows the result on the absorption spectrum $\rm{Im}(\rho_{21}^A)$ as a function of $\Delta_c$ through solving the master equation~(\ref{Master}) numerically. We see that $\rm{Im}(\rho_{21}^A)$ has an EIT-like pattern, where a transparent window (called {\it FIT transparent window} hereafter) in the absorption spectrum opens. For weak inter-channel interaction (i.e. small $V_{AB}$), Im($\rho_{21}^A$) displays only a single absorption peak, while for stronger interaction (i.e. larger $V_{AB}$) two absorption peaks occur [see Fig.~\ref{FIT_Windows}(c1)]. Concretely, in the region of large $V_{AB}$, one peak (the lower one) locates at $\Delta_c=0$ and the other (the higher one) locates at $\Delta_c= -V_{AB}$. The width of the two peaks is $\Delta E$ [see Fig.~\ref{FIT_Windows}(b)]. Our calculating result also shows that $\rho_{21}^A$ strongly correlates to Rydberg populations  $\rho_{33}^{A}$ and $\rho_{33}^{B}$.

In Fig.~\ref{FIT_Windows}(c1), one can observe clearly that there is a crossover from one peak to two ones for $\rm{Im}(\rho_{21}^A)$ through the adjustment of $V_{AB}$, and the transparent window becomes wider and deeper as inter-channel interaction $V_{AB}$ increases. This fact tells us that the splitting of the probe-field absorption spectrum (i.e. FIT) is facilitated by the contribution of the inter-channel interaction $V_{AB}$.
For illustrating this point further, the left part of Fig.~\ref{FIT_Windows}(a)
[which corresponds to the point indicated by the small green circle in Fig.~\ref{FIT_Windows}(b)]
gives the level diagram and excitation scheme between the Ryder blockade state $|1_A 3_B\rangle$ (occurring for $\Delta_c=0$) to the facilitated state $|3_A3_B\rangle$ (occurring for $\Delta_c=-V_{AB}$) of the two-atom model; the solid gray circles (solid blue circles) denote ground-state (Rydberg state) atoms. The right part of the figure is a simplified representation of the excitation scheme showing in the left part. Although the transition of the target atom $A$ to the Rydberg state acquires an energy shift because of the inter-channel interaction $V_{AB}$, by sweeping the detuning $\Delta_c$ in the channel $B$ the system can be driven from the Rydberg blockade state $|1_A 3_B\rangle$ to the facilitated state $|3_A3_B\rangle$,
where the channels $A$ and $B$ are simultaneously excited to the Rydberg state near at $\Delta_c=-V_{AB}$ [see Fig.~\ref{FIT_Windows}(b) and (c2)].
Since the positions of the excitation blockade and the facilitation overlaps well with the peaks of $\rm{Im}(\rho_{21}^A)$, the FIT transparency window between the two strong absorption peaks is thus induced by the both effects of the excitation blockade and the facilitation.

To verify the behavior of the atomic coherence and the FIT phenomenon in a direct way, we have solved the Heisenberg equations of motion of the system.
The calculation shows that $\rho_{21}^A$ is given by the following form
\begin{eqnarray}\label{Correlation}
\rho_{21}^A\simeq -\frac{2V_{AB}\,\rho_{31,33}^{AB}}{\Omega},
\end{eqnarray}
where $\rho_{31,33}^{AB}\equiv \langle\hat{\sigma}_{13}^{A}\hat{\sigma}_{33}^{B}\rangle$ is two-body DM element (two-body correlator) [see Appendix~\ref{ap2} for details].

The result predicted by using this approach is plotted in Fig.~\ref{FIT_Windows}(d) by a dotted red line. One sees that the analytical (\ref{Correlation}) captures well with the one by solving the master equation (ME) numerically (given by the solid black line). One notes that the variance $O_{AB}=\langle\hat{\sigma}_{13}^{A}\hat{\sigma}_{33}^{B}
\rangle-\langle\hat{\sigma}_{13}^{A}\rangle\langle\hat{\sigma}_{33}^{B}\rangle
=\rho_{31,33}^{AB}-\rho_{31}^{A}\rho_{33}^{B}$ exhibits the similar double peak structure of the FIT.
In the above calculations, system parameters used are $\Omega=\Omega_c=5\Gamma_{12}^A$, $\Omega_p=0.5\Gamma_{12}^A$, $\Gamma_{23}^A=10^{-3}\Gamma_{12}^A$, $\Gamma_{13}^B=\Gamma_{12}^A$, and $V_{AB}=15\Gamma_{12}^A$ in panels (b) and (d). Based on the results described in Fig.~\ref{FIT_Windows}, we conclude that FIT is mainly contributed by the inter-channel quantum correlation, contributed by the Rydberg interaction between the atom $A$ and the atom $B$. It should be stressed that such results cannot be predicted by using the approach of mean field theory~\cite{Coherent_Schempp_2010, pritchard_cooperative_2010}. This indicates that the two-body correlation could be effectively mapped to the atomic coherence $\rho_{21}^A$, providing a method to detect Rydberg facilitation directly from the optical absorption spectrum, which is the key finding of this work.

\subsection{Excitation pathways identified through dressed-state approach and generation to multiple target channels}\label{sec_dressed}

Even when the system involves only two atoms, it is difficult to solve the master equation analytically (e.g. $36$ coupled  differential equations are involved). Motivated by the dressed-state theory of conventional EIT~\cite{Fleischhauer_Fleischhauer_2005}, here we develop an analytical approach  of dressed state  to identify the excitation pathways of the FIT through the facilitated many-body states.

\begin{figure}[htb]
\centering
\includegraphics[width=0.5\textwidth]{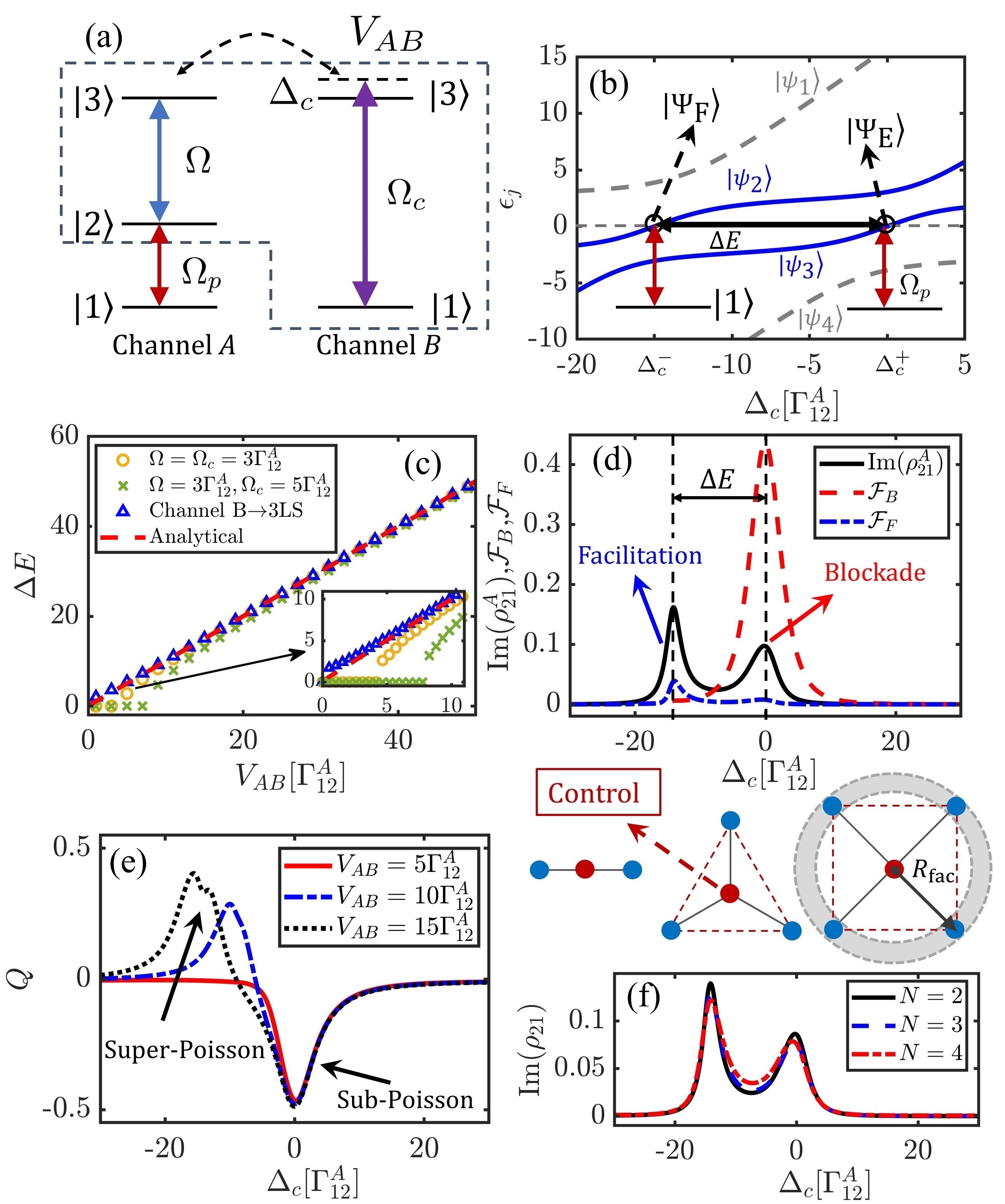}\\
\caption{\footnotesize\textbf{Dressed-state picture and the scaling of the FIT peak.} (a) Interaction-dressed subspace spanned by the basis $\{|2_A\rangle, |3_A\rangle\}\otimes\{|1_B\rangle, |3_B\rangle\}$, indicated by the dashed gray box.
(b) Dressed eigen values $\epsilon_j$ as functions of $\Delta_c$ for $V_{AB}=15\Gamma_{12}^A$; corresponding eigenstates $\{|\psi_j\rangle; j=1,4\}$ (dashed gray lines) and  $\{|\psi_j\rangle; j=2,3\}$  (solid blue lines) are also shown.
$\Omega_p$ can resonantly drive the transition $|1_A\rangle\rightarrow|\Psi_{F}\rangle$  ($|1_A\rangle\rightarrow|\Psi_{E}\rangle$) at $\Delta_c=\Delta_c^{-}$ ($\Delta_c=\Delta_c^{+}$), where $|\Psi_{E}\rangle$ ($|\Psi_{F}\rangle$) is the blockade (facilitation) state (see the Appendix~\ref{ap3}).
(c) Energy separation $\Delta E$ [described also in Fig.~\ref{FIT_Windows}(a)] as a function $V_{AB}$ for $\Omega_c=3\Gamma_{12}^A$ (yellow circles) and  $\Omega_c=5\Gamma_{12}^A$ (green crosses) with fixed $\Omega=3\Gamma_{12}^A$ and $\Gamma_{13}^B=0.1\Gamma_{12}^A$, obtained by numerically solving the master equation~(\ref{Master}).
Triangles: the result for the channel $B$ is driven by a three level  scheme (3LS).
Inset: details of $\Delta E$ in the different parameter regimes for small $V_{AB}$.
Dashed red line: the line $\Delta E=V_{AB}$, which matches well with the numerical data for larger $V_{AB}$.
(d) Fidelities ${\cal F}_{B}$ (dashed red line) and  ${\cal F}_{F}$ (dash-dotted blue line) as functions of $\Delta_c$.  Maximal fidelities overlap with the doublet peaks of the absorption spectrum $\rm{Im}(\rho_{21}^A)$ (solid black line).
(e) Mandel $Q$ factor as a function of $\Delta_c$ for different $V_{AB}$, which
is enhanced drastically at $\Delta_c=\Delta_c^{\pm}$
for large  $V_{AB}$.
(f) Suggested model with multiple target channels, in which the control channel (solid red circle) locates at the center of a ring (with radius $R_{\rm fac}$) where the target channels (solid blue circles) locate. Such a setting can facilitates the excitation of $N$-target channels (solid blue circles) within a shell (gray region). Shown in the
box are lines of Im($\rho_{21}$) for $N=2, 3, 4$, indicating that FIT is supported for such a system.} \label{Scaling}
\end{figure}

We consider the interaction-dressed subspace spanned by the tensor product of two single-atom basis, i.e. $\{|2_A\rangle, |3_A\rangle\}\otimes\{|1_B\rangle, |3_B\rangle\}$, indicated by the dashed gray box in Fig.~\ref{Scaling}(a). The
subspace is coupled by the ground state $|1_A\rangle$ of the channel $A$ through the probe field $\Omega_p$.
For simplicity and for obtaining explicit analytical results, we neglect the small damping (i.e. the spontaneous emission and dephasing). The eigenspectra $\epsilon_j$ of the subspace can be obtained via the diagonalization of the system Hamiltonian (see Appendix~\ref{ap3}) expanded in this subspace, with corresponding set of eigen basis given by $\{\,|\psi_j\rangle$; $j=1,2,3,4\}$.

To highlight the roles played by the control channel $B$, we make a calculation especially on the eigenspectra and eigenfunctions. Shown in Fig.~\ref{Scaling}(b) is the result for the eigenvalues $\epsilon_j$ as functions of $\Delta_c$ for $V_{AB}=15\Gamma_{12}^A$. Corresponding eigenstates $\{|\psi_j\rangle; j=1,4\}$ and $\{|\psi_j\rangle; j=2,3\}$ are given by the dashed gray and  solid blue lines, respectively. We see that the prove field $\Omega_p$ can resonantly drive the transition $|1_A\rangle\rightarrow|\Psi_{F}\rangle$  ($|1_A\rangle\rightarrow|\Psi_{E}\rangle$) at $\Delta_c=\Delta_c^{-}$ ($\Delta_c=\Delta_c^{+}$), where $|\Psi_{E}\rangle$ ($|\Psi_{F}\rangle$) is the blockade (facilitation) state. For more details, see the discussion given in the Appendix~\ref{ap3}.

From the figure, we see that, when $\epsilon_j=0$, the probe field $\Omega_p$ can resonantly drive the transition from $|1_A\rangle$ to the dressed subspace at the particular detuning
$\Delta_c^{\pm}=-V_{AB}/2\pm\sqrt{V_{AB}^2+\left(\Omega-\Omega_c^2/\Omega\right)^2}/2$,
which locates the double peaks of the absorption spectrum Im($\rho_{21}^A$) [predicted also in Fig.~\ref{FIT_Windows}(b)], with energy separation $\Delta E=\Delta_c^+-\Delta_c^-$. In the strong interaction dominant regime (i.e. $|V_{AB}|>|\Omega|,\, |\Omega_c|\gg |\Omega - \Omega_c^2/\Omega|$), one can obtain $\Delta E\simeq V_{AB}$. This linear dependence on $V_{AB}$ for the distance between the two absorption peaks can also be seen in the numerical data given by Fig.~\ref{Scaling}(c). Deviations for smaller $V_{AB}$ attribute to spontaneous emission, dephasing of the atomic coherence and light induced shifts, which are included in the numerical simulation. In Fig.~\ref{Scaling}(c), the result for the case when the control channel described by three-level (3L) model (e.g. two-photon Raman scheme) is also shown. One sees that the energy separation $\Delta E$ as a function $V_{AB}$ also have a linear dependence for large $V_{AB}$.

Now lets turn to investigating the eigenstates of the system, which, in the interaction dominant region, are given by
\begin{eqnarray}
|\Psi_E\rangle&=&\frac{1}{\sqrt{2}}(|3_A1_B\rangle-|2_A3_B\rangle),\,\,\,\,{\rm for}\,\, \Delta_{c}=\Delta_c^{+},\\
|\Psi_F\rangle&=&\frac{1}{\sqrt{2}}(|3_A3_B\rangle-|2_A1_B\rangle),\,\,\,\,{\rm for}\,\, \Delta_{c}=\Delta_c^{-},
\end{eqnarray}
where $|\Psi_{E(F)}\rangle$ are Rydberg entangled and facilitated Bell states. The excitation of these states provides two pathways. When the inter-channel interaction is strong, they interfere destructively and lead to the FIT.
However, due to the weak probe field and the large spontaneous emission at the intermediate level $|2_A\rangle$, the entangled state $|\Psi_E\rangle$ decays to the blockade state $|\Psi_B\rangle=|1_A3_B\rangle$, which partially destroys the transparency between the two absorption peaks, consisted with the result given in  Fig.~\ref{FIT_Windows}.

To verify the validity of the dressed-state approach in the presence of dissipation, we quantify the difference between states $\hat{\rho}_{B(F)}=|\Psi_{B(F)}\rangle\langle\Psi_{B(F)}|$ and the steady state $\hat{\rho}_s$ from master equation simulations with fidelity ${\cal F}_{B(F)}=({\rm Tr}|\sqrt{\hat{\rho}_{B(F)}}\sqrt{\hat{\rho}_s}|)^2$. When sweeping $\Delta_c$, ${\cal F}_{B(F)}$ indeed displays maxima at $\Delta_c^\pm$ [see Fig.~\ref{Scaling}(d)]. This indicates that the dressed-states can be used to characterize the Rydberg blockade and facilitation in the mixed-state system.

We emphasize that two-body quantum correlations are important to the FIT. We further quantify the inter-channel correlation by the Mandel $Q$ factor defined as $Q=\langle(\Delta \hat{n}_{33})^2\rangle/\langle\hat{n}_{33}\rangle-1$ with $\hat{n}_{33}=\hat{\sigma}_{33}^{A}+\hat{\sigma}_{33}^{B}$. As illustrated in Fig.~\ref{Scaling}(e), Mandel $Q$ factor is enhanced drastically at $\Delta_c=\Delta_c^{\pm}$, and exhibits sub-Poissonian counting statistics
with $Q<0$ around $\Delta_c=\Delta_c^{+}$, and the super-Poissonian processes characterized by $Q>0$ around $\Delta_c=\Delta_c^{-}$. This is an unique property for such a system, where exotic statistics characteristics can be probed via tuning laser parameters~\cite{Statistics_Schempp_2014,Valado_Experimental_2016}.

The dual channel model discussed above can be extended to the situations with multiple target channels. One possible generation is sketched in Fig.~\ref{Scaling}(f), in which the control channel (denoted by the solid red circle) locates at the center of a ring (with radius $R_{\rm fac}$) where the target channels (denoted by solid blue circles) locate. Such a setting can facilitates the excitation of $N$-target channels (represented by solid blue circles) within a shell (represented by the gray region)~\cite{Out_Lesanovsky_2014}. The lower part of the figure (i.e. the
box) shows the lines of the probe-field absorption spectrum Im($\rho_{21}$) for the target-channel number $N=2, 3, 4$, which indicates that FIT character still persists with multiple target channels.

In our simulation, interactions between different target channels have been taken into account. Due to the strong interaction contributed from the control channel, the target channels are fully blockade or facilitated, and thus the target-target (TT) interaction plays a minor role on the FIT~\cite{ Marcuzzi_Facilitation_2017,Signatures_Helmrich_2020}. To demonstrate this point further, we have constructed a three-atom FIT model (in which one control atom and two target atoms are considered) and investigated its FIT behavior. The results shows that the TT interaction have a small effect on the FIT in the system; for more detail, see Appendix~\ref{ap4}.

\section{Facilitated single-photon switch protocol}\label{sec_switch}

The above analysis shows that the target channel $A$ can be changed from opaque to transparent by tuning $\Delta_c$. Such tunability allows us to design optical switch via FIT by using contactless atom-atom interactions. In the rest of the work, we consider the propagation dynamics of quantized probe fields in high density atom gases. We will show that a facilitated single-photon switch can be realized in the dual-channel system suggested above.

\subsection{Optical susceptibility of the probe field}
Before examining quantum dynamics of the single-photon switch, we first analyze the polarization of the probe field ${\cal E}_p$, defined by ${P}_p\equiv\epsilon_0\chi_p {{\cal E}}_p$. The susceptibility $\chi_p$ is, in general, of a nonlinear function of ${\cal E}_p$ if the probe-field intensity is high. The relation between the optical susceptibility of the probe field and the one-body DM elements can be built with the formula
\begin{equation}\label{susceptibility}
\chi_p=\frac{\hbar\varepsilon_0}{{\cal N}_A \mu_{21}^2\Omega_p}\, \rho_{21}^A,
\end{equation}
where ${\cal N}_A$ is the atomic density of the ensemble $A$ (assumed to be homogeneous). In our setting, $\chi_p$ depends not only on the parameters of the ensemble $A$ (e.g, $\Omega_p$, $\Omega$), but also on those of the ensemble $B$ (e.g., $\Omega_c$, $\Delta_c$).
\begin{figure}[htb]
	\centering
	\includegraphics[width=0.45\textwidth]{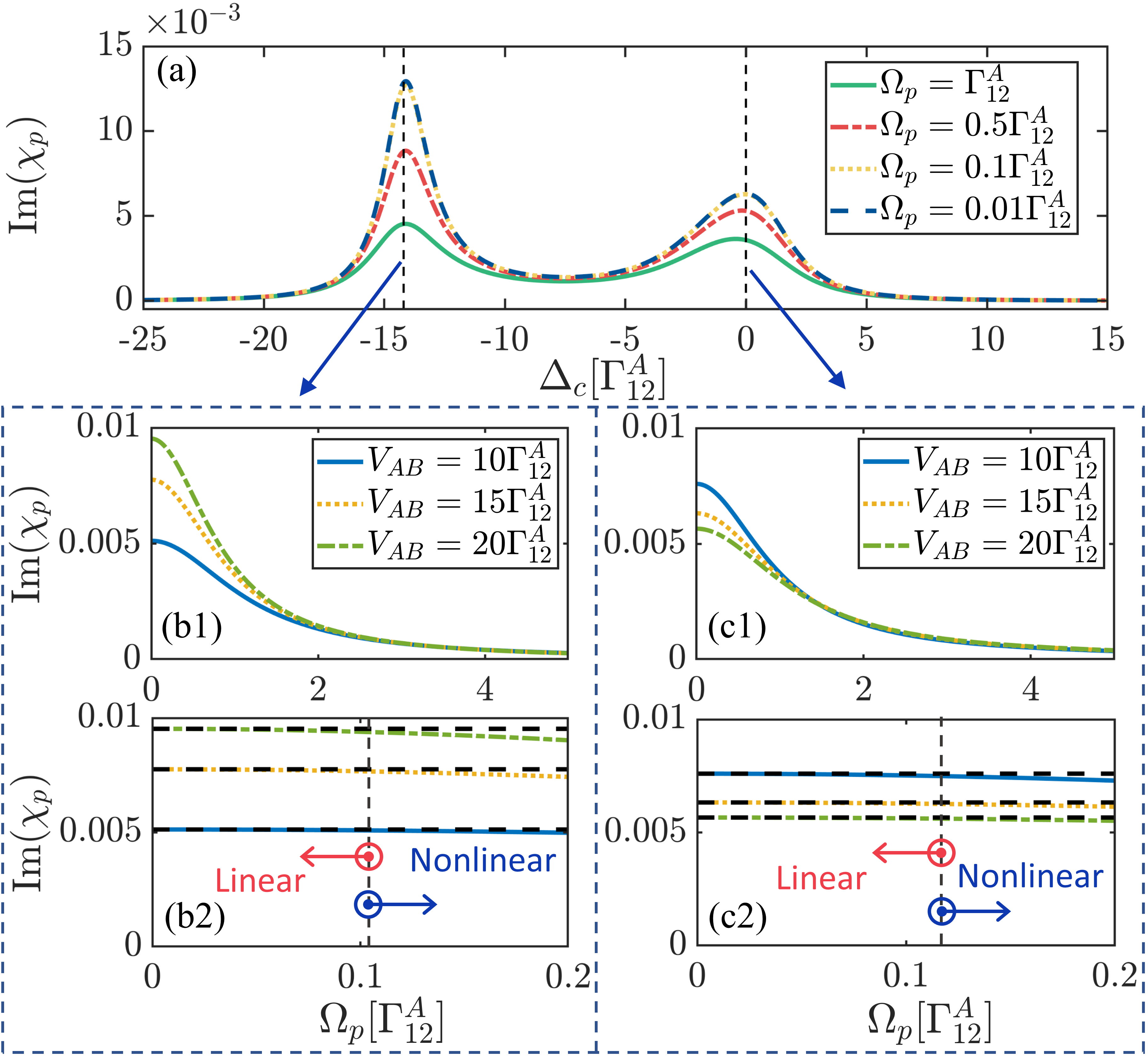}\\
	\caption{\footnotesize\textbf{Optical susceptibility.} (a) Im($\chi_p$) as a function of $\Delta_c$ (for $V_{AB}=15\Gamma_{12}^A$) when the Rabi frequency of the probe field takes values of $0.01\Gamma_{12}^A, 0.1\Gamma_{12}^A, 0.5\Gamma_{12}^A,$ and $\Gamma_{12}^A$, respectively.
(b1) Im($\chi_p$) as a function of $\Omega_p$ and $V_{AB}$ for $\Delta_c=-V_{AB}$ [corresponding to the left peak of panel (a)].
(b2) The same as (b1) but for small values of $\Omega_p$.
(c1) and (c2): The same as (b1) and (b2) but with $\Delta_c=0$ [corresponding to the right peak of panel (a)].
The result shows that when decreasing $\Omega_p$, Im($\chi_p$) saturates for $\Omega_p<0.1\Gamma_{12}^A$, such that $\chi_p$ behaves as a linear susceptibility,  independent of $\Omega_p$.} \label{linear_susceptibility}
\end{figure}

Figure~\ref{linear_susceptibility}(a) shows the imaginary part of the susceptibility Im($\chi_p$) as a function of Rabi frequency of the probe field, $\Omega_p$, with fixed $V_{AB}=15\Gamma_{12}^A$. The dashed blue, dotted yellow, dot-dashed red, and solid green lines are for $\Omega_p=0.01\Gamma_{12}^A, 0.1\Gamma_{12}^A, 0.5\Gamma_{12}^A,$ and $\Gamma_{12}^A$,
respectively. Illustrated in panels (b1) is Im($\chi_p$) as a function of $\Omega_p$ and $V_{AB}$ for $\Delta_c=-V_{AB}$, which corresponds to the left peak of panel (a); panel (b2) is same as panel (b1) but for small values of $\Omega_p$. Panels (c1) and (c2) are the same as (b1) and (b2) but with $\Delta_c=0$ [corresponding to the right peak of panel (a)].
The results are obtained by solving the Heisenberg equations of the system.
From the figure we see that, as the probe Rabi frequency is decreased, there is a dramatic enhancement of Im($\chi_p$) around $\Delta_c=0$ and $-V_{AB}$, manifested by the two FIT peaks in the figure. Moreover, there is no broadening or detuning of the FIT, and the susceptibility saturates when $\Omega_p < 0.1\Gamma_{12}^A$, for which Im($\chi_p$) does not change as $\Omega_p$ is decreased.
The absence of the broadening or shift is vital for the detection of Rydberg facilitation from the optical absorption spectrum. When the probe field  is weak (e.g. at the level of single photons), $\chi_p$ becomes linear, i.e. its value is largely independent of ${\cal E}_p$ in both the facilitation and blockade region [see Fig.~\ref{linear_susceptibility}(b2) and (c2)]. Outside the linear regime, one can achieve a strong Rydberg nonlinearity in such a system, which is a topic deserving explorations but beyond the scope of the present work.

\subsection{Facilitated single-photon switch}

In the following, we will focus on the linear regime and demonstrate how to realize a single-photon switch with the FIT predicted above. To this end, we consider an elongated atomic gas and study propagation of the probe field along the $z$ direction, while neglecting diffraction effect, which is valid for the case of short propagation distance or the atomic gas is prepared in a trap with a small transverse cross section~\cite{Single_Gorniaczyk_2014,Busche_Contactless_2017,Two_Chen_2021}. Treating the probe field quantum mechanically, the dynamics of its propagation is governed by the
Hamiltonian $\hat{H}_p = -(c/L)\int {dz\hat{{\cal E}}_p^\dagger(z,t)i\partial_z\hat{{\cal E}}_p(z,t)}$. Here, $\hat{{\cal E}}_p(z,t)$ is the slowly varying annihilation operator of probe photons,  satisfying the equal-time commutation relation $[\hat{{\cal E}}_p(z,t), \hat{{\cal E}}_p^\dagger(z^\prime,t)]=L\delta(z-z^\prime)$, with $L$ the quantization length along the $z$ axis.
By applying the continuous density approximation~\cite{fleischhauer_quantum_2002} 
the Hamiltonian of the system reads
\begin{subequations}\label{Hami}
\begin{eqnarray}
\hat{H}_{A} &=& -\frac{N_A}{L}\int {dz} \left[\frac{g\hat{{\cal E}}_p(z, t)}{2}\hat{\sigma}_{21}^A(z, t)+\frac{\Omega}{2}\hat{\sigma}_{32}^A(z, t)+{\rm H.c.}\right],\nonumber\\\\
\hat{H}_{B} &=& -\frac{N_B}{L}\int {dz} \left[\frac{\Omega_{c}}{2}\hat{\sigma}_{31}^B(z, t)-\Delta_c\hat{\sigma}_{33}^B(z, t)+{\rm H.c.}\right],
\end{eqnarray}
\end{subequations}
here $N_A$ and $N_A$ are respectively atomic numbers in the channels $A$ and $B$,
$\hat{\sigma}_{ab}^\mu(z)$
satisfies the canonical commutation relation $[\hat{\sigma}_{ab}^\mu(z), \hat{\sigma}_{cd}^\mu(z')]=\delta(z-z^\prime)[\delta_{bc}\hat{\sigma}_{ad}^\mu(z)
-\delta_{da}\hat{\sigma}_{cb}^\mu(z)]L/N_\mu$.
The Hamiltonian describing the inter-channel interaction is given by
\begin{eqnarray}\label{Hami1}
H_{\rm int}=\frac{N_A}{L}\sum_j\int {dz}V(z-z_j)\hat{\sigma}_{33}^{A}(z, t)\hat{\sigma}_{33}^{B}(z_j, t),
\end{eqnarray}
where the index $j$ labels the respective Rydberg excitation in the channel $B$~\cite{Weibin_Coherence_photon_switch2015}. We assume that the atoms within each channel are trapped within a small spatial interval $\delta r\ll d$; the atom-atom interaction is described by the potential  $V(z-z_j)\simeq -C_6/[(z-z_j)^2+d^2]^3$ with $V_{AB}=-C_6/d^6$~\cite{Busche_Contactless_2017}.

Suppose that the quantum state of the single-photon input probe field has the form $|\Psi_P\rangle=(1/L)\int {dz}E_p(z,t)\hat{{\cal E}}_p^\dagger(z,t)|0\rangle$, where $E_p(z, t)$ is the single-photon probability amplitude. Based on the Maxwell equation for $\hat{\cal E}_p(z,t)$, one can derive the evolution equation for $E_p(z,t)$~\cite{Gorshkov_Photon_2007,Chong_Photon_2021}, which reads
\begin{eqnarray}\label{Max}
&& \left(\frac{\partial}{\partial t}+c\frac{\partial}{\partial z}\right)E_p(z, t)= ig\,N_A \rho_{21}^A(z, t)/2.
\end{eqnarray}
It can be solved numerically in conjunction with the Bloch equations for the atomic DM elements $\rho_{ab}^{A}$ and $\rho_{ab}^{B}$. The transmission $T$ of
the single-photon probe field, used to quantify efficiency of the photon switch, can be obtained at $z=L$, i.e. $T=\mathcal{I}_s(L)$, where $\mathcal{I}_s(z)=|E_p(z)|^2/|E_p(0)|^2$ is the relative intensity of the photon field at position $z$.

\begin{figure}[htb]
\centering
\includegraphics[width=0.5\textwidth]{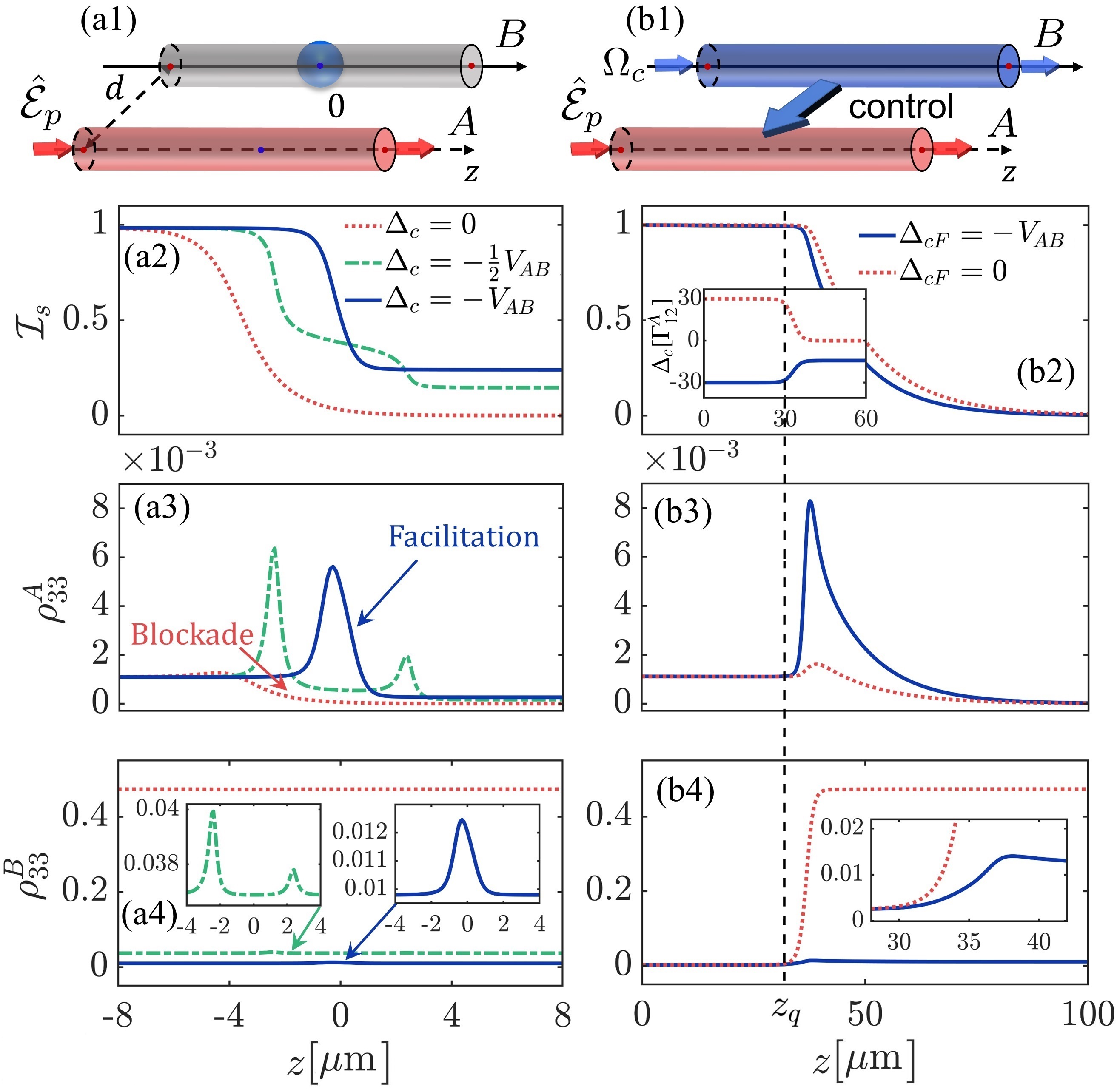}\\
\caption{\footnotesize\textbf{Facilitation enabled single-photon switch.}
(a)~The control channel $B$ is driven with the excitation localized at $z=0$ (a1).  In the vicinity of $z=0$, $\rho_{33}^{A}$ is blockaded for $\Delta_c=0$ or facilitated for $\Delta_c=-V_{AB}$ ($\Delta_c=-V_{AB}/2$ is an intermediate case). They both result in a strong photon scattering, as can be seen from the decreasing relative intensity $\mathcal{I}_s$ (a2).
Shown in (a3) and (a4) are $\rho_{33}^{A}$ and $\rho_{33}^{B}$ as functions of $z$ for $\Delta_c=0$ (red dashed line), $-V_{AB}/2$ (green dash-dotted line), and  $-V_{AB}$ (blue solid line), respectively.
(b) When the channel $B$ is dynamically driven (b1), one can change the control detuning $\Delta_{c}$ from a far-detuned value to $-V_{AB}$ (or $0$) to launch facilitated (blockade) single-photon switch, where the relative intensity of the probe field quickly decreases to zero (b2).  Shown in (b3) and (b4) are $\rho_{33}^{A}$ and $\rho_{33}^{B}$ as functions of $z$ for $\Delta_{cF}=0$ (red dashed line), and  $-V_{AB}$ (blue solid line), respectively.
In the simulation, system parameters  $\Omega=\Omega_c=3\Gamma_{12}^A$, $V_{AB}=15\Gamma_{12}^A$, and $\mathcal{N}_A=3\times10^{12}{\rm cm}^{-3}$ are used.
See text for more details.
}
\label{Switching}
\end{figure}

To understand the propagation dynamics of the probe photon, lets first consider a simple scenario in which the Rydberg excitation in the control channel $B$ is localized at $z=0$. Faraway from $z=0$, the photon field can propagate under perfectly EIT condition in the channel $A$ [where the excitation scheme is the one of an exact ladder-type EIT; see Fig.~\ref{Model}(a)], as illustrated in Fig.~\ref{Switching}(a1).
However, in close proximity to $z=0$ (noticing that $V_{AB}$ varies with $z$ and reaches its maximal value $V_{AB}$ at $z=0$), the channel $B$ inhibits ($\Delta_c=0$) or facilitates ($\Delta_c=-V_{AB}$) the Rydberg excitation in the channel $A$, and causes strong photon loss from the medium; see Fig.~\ref{Switching}(a2), where $\mathcal{I}_s$ as a function of $z$ is plotted
for $\Delta_c=0$, $\Delta_c=-V_{AB}/2$, and $\Delta_c=-V_{AB}$, respectively.
The former (i.e. the result for $\Delta_c=0$) is the phenomenon of Rydberg blockade, which has been studied in Ref.~\cite{Single_Baur_2014, Single_Gorniaczyk_2014, Weibin_Coherence_photon_switch2015}; the latter (i.e. the result for $\Delta_c=-V_{AB}$) behaves as a {\it facilitated photon switch} where the probability of the Rydberg excitation in the both channels is relatively large. This can be seen clearly in Fig.~\ref{Switching}(a3) and (a4), where
$\rho_{33}^{A}$ and $\rho_{33}^{B}$ as functions of $z$. When passing from the interaction region, the photon field  propagates freely again under the EIT condition where the photon scattering is negligible. Therefore, the reduced transmission is caused by the facilitation induced photon scattering in the vicinity of $z=0$.

For an ideal photon switch, the photon field should be completely blocked from transmission (i.e. $T=0$). To achieve this, we consider the light field in the control channel co-propagate with the probe field in the target channel, depicted in Fig.~\ref{Switching}(b1). To demonstrate the controllability of the photon switch, the detuning $\Delta_c$ is chosen to be time-dependent but varies slowly, which has the from of $\Delta_c=\Delta_{c0}+(\Delta_{c0}-\Delta_{cF})[{\rm tanh}(1-(z-z_q)/z_s)-1]/2$ [see the inset of Fig.~\ref{Switching}(b2)]. Initially, $\Omega_c$ is far detuned before the photon field travels to $z=z_q$. Then it is rapidly changed to $\Delta_{cF}\simeq -V_{AB}$ (or $0$). In response to such changes, the system immediately evolves from a non-interacting dark states into an absorbing facilitated state (or blockade state) [see the behavior of $\rho_{33}^{A(B)}$ shown in Fig.~\ref{Switching}(b3)-(b4)]. The transmission of the photon field quickly decreases from almost $100\%$ to zero within a short propagation distance. Consequently, for a medium of less than 100 $\mu$m, the photon field transmission can be made completely vanish, which is promising for designing highly efficient and controllable single-photon switches.

\subsection{Discussion on delocalized control-atom excitation}

In the above consideration, we have assumed the spatial distribution of each atom ensemble is small, the excitation of the control atom is treated to be localized at a fixed position in the channel $B$, such that the interaction is described by the simple form of $V_{AB}=-C_6/d^6$. In reality, atoms in the ensemble spread in space~\cite{Busche_Contactless_2017}. For practical realization of the FIT and photon switch predicted here, the spatial dependence of the atom-atom interaction needs to be taken into account.

\begin{figure}
\centering
\includegraphics[width=0.5\textwidth]{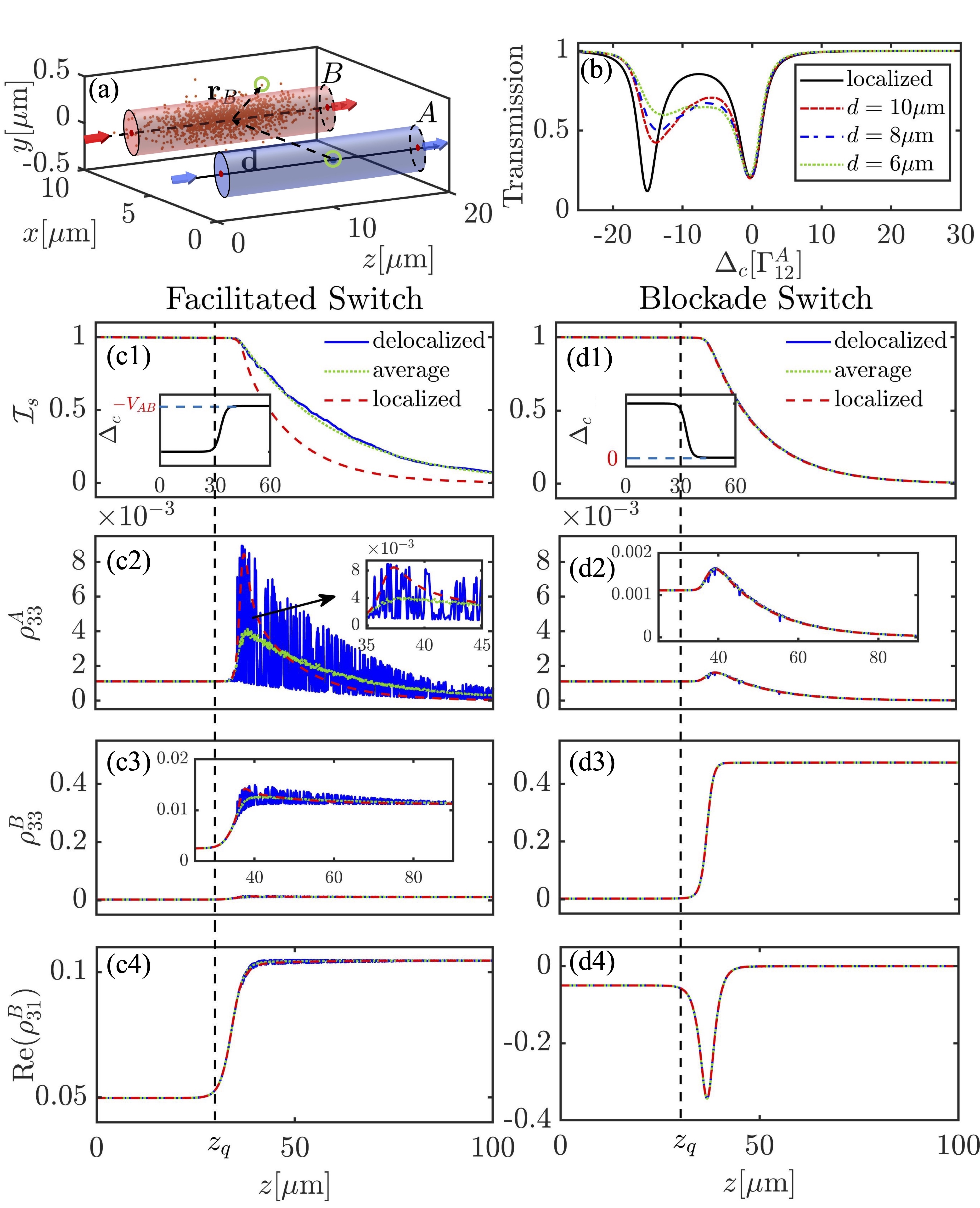}\\
\caption{\footnotesize \textbf{FIT with delocalized control-atom excitation}.
(a)~Schematic of the delocalized control-atom excitation in the channel $B$. The delocalization modifies
the atom-atom interaction form from $V_{AB}=-C_6/d^6$ to  $V_{AB}=-C_6/|\mathbf{d}+\mathbf{r}_B|^6$, here  $d=|\mathbf{d}|$ is the center-of-mass separation between the $A$ and $B$ channels and  $r_B=|\mathbf{r}_B|$ is the relative variation of the control excitation.
(b) Photon transmission $T$ as a function of $\Delta_c$ for different $d$, but with fixed $V_{AB}(r_B=0)=15\Gamma_{12}^A$, when the photon propagates to distance  $L=20\,{\rm \mu m}$.
Solid black line: the result of the ideal FIT (i.e. the case for localized control-atom excitation).
(c) is similar to Fig.~\ref{Switching}(b). To launch facilitated (blockade) switch, here $\Delta_c$ is changed from the far-detuned value to $-V_{AB}$ (or $0$).
(c1)-(c3) show the relative intensity of the probe field $I_s$ and the Rydberg population $\rho_{33}^{A(B)}$ in the facilitated regime; (d1)-(d3) are similar to (c1)-(c3) but for the blockade regime. Blue curves in these figures are results of single trajectory simulation in the delocalized regime. Their mean values are calculated by casting 300 trajectories for different initial delocalized distribution (see green-dotted line).
The coherence of the control atom is modified  during the operation of the photon switch, manifested by the real Re($\rho_{31}^B$) of the atomic coherence [see panel (c4) and (d4)]. System parameters used in the simulation are $\Omega=\Omega_c=3\Gamma_{12}^A$ and $\mathcal{N}_A=3\times10^{12}{\rm cm}^{-3}$.
}
\label{delocalized}
\end{figure}

When there is a delocalization of the control-atom excitation over the channel $B$, the inter-channel interaction takes the form  $V_{AB}=-C_6/|\mathbf{d}+\mathbf{r}_B|^6$. Here $d=|\mathbf{d}|$ is the center-of-mass (COM) separation between the channels $A$ and $B$, and  $r_B=|\mathbf{r}_B|$ is the distance of the control excitation relative to
the COM of the  channel $B$; see Fig.~\ref{delocalized}(a).
Under such a consideration, we can obtain $\mathbf{r}_B$-dependent $\rho_{21}^A$, which reads
\begin{eqnarray}\label{rb_rho}
\rho_{21}^A(\mathbf{r}_B)\simeq -\frac{2V_{AB}(\mathbf{r}_B)\,\rho_{31,33}^{AB}(\mathbf{r}_B)}{\Omega}.
\end{eqnarray}
Then optical susceptibility of the probe field is given by
\begin{equation}\label{susceptibility}
\chi_p=\int d\mathbf{r}_B f(\mathbf{r}_B)\frac{\hbar\varepsilon_0}{{\cal N}_A \mu_{21}^2\Omega_p}\, \rho_{21}^A(\mathbf{r}_B),
\end{equation}
where $f(\mathbf{r}_B)$ is a normalized distribution function describing the spatial delocalization of the control-atom excitation in the channel $B$. For simplicity, we assume $f(\mathbf{r}_B)$ takes the form of the Gaussian distribution, i.e.  $f(\mathbf{r}_B)=1/(\sqrt{\pi}\sigma){\rm exp}[-(\mathbf{r}_B/\sigma)^2]$), with $\sigma$ being the distribution width.
The delocalization degree can be characterized by using  parameter ${\rm max}(r_B)/d$.

Fig.~\ref{delocalized}(a) shows the schematic of the delocalized control-atom excitation in the channel $B$ with a three-dimensional Gaussian-shaped distribution.
To get a quantitative result of the behavior of single-photon switch in the presence of the delocalized control-atom excitation, a numerical simulation is carried out by using a combination of the four-order Runge-Kutta and finite difference methods to solve the corresponding coupled  Maxwell-Bloch equations.
In the simulation, we average the transmission $T$ over different sites until the result converges.

Shown in Fig.~\ref{delocalized}(b) is the transmission $T$ as a function of $\Delta_c$ for different separation $d$, but with  $V_{AB}(r_B=0)=15\Gamma_{12}^A$. To obtain such an interaction strength, it requires $C_6=2.68\times10^4{\rm GHz\cdot\mu m}^6$ (with state $|93S\rangle$) for $d=6{\rm \mu m}$ and $C_6=5.77\times10^5{\rm GHz\cdot\mu m}^6$ (with state $|122S\rangle$) for $d=10{\rm \mu m}$  in Rb atom.
We can see that, in the presence of the delocalized control-atom excitation, the FIT can still be observed. However, in comparison with the case of the localization regime (i.e., the regime with no position fluctuation of the control-atom excitation), there has a slight shift on the absorption peak located at $\Delta_c= -V_{AB}$ (shown by the black solid line). In addition, in the delocalization regime, an inhomogeneous broadening for the FIT peak occurs; as $d$ decreases, the delocalization degree of the  control-atom excitation increases, and the FIT peak becomes smaller. Consequently, to realize a robust FIT, the system must have strong Rydberg interactions, larger separation between the two channels, and tightly-focused  control field.


To test the protocol of single-photon switch, we have conducted a simulation similar to that shown in Fig.~\ref{Switching}(b), but with delocalized control-atom excitation here.  Fig.~\ref{delocalized}(c1) shows the results of the relative probe-field intensity ${\cal I}_s$ as a function of $z$, for regimes of the delocalization (blue solid line) and localization (red dashed  line) are given. We see that ${\cal I}_s$ decreases slower in the delocalization regime but still approaches to zero within a short propagation distance.
Drawn in Fig.~\ref{delocalized}(c2) [Fig.~\ref{delocalized}(c3)] is the behavior of the Rydberg population $\rho_{33}^{A}$ ($\rho_{33}^{B}$). One sees that, due to the delocalized excitation, both $\rho_{33}^{A}$ and  $\rho_{33}^{B}$ exhibit large fluctuations;
however, their envelopes are still facilitated, and captured well with the corresponding localized curves, respectively.  To see the average behaviors of ${\cal I}_s$ and $\rho_{33}^{A(B)}$ in the presence of the fluctuations,
in the simulation 300 trajectories for different initial delocalization distributions are considered,  and these quantities are evaluated by calculating their mean values. The results of such an average are given by the green-dotted  lines in the figure.

Shown in Fig.~\ref{delocalized}(d1)-(d3) are also for the quantities ${\cal I}_s$ and $\rho_{33}^{A(B)}$, but for the case of the blockade regime. Blue curves in these figures are results of single trajectory simulation in the delocalization regime. Their mean values are also calculated by casting 300 trajectories for different initial delocalized distributions (give by green-dotted lines).  We see that the quality of the blockade switch is not to be significantly affected by the delocalized distribution of the control-atom excitation.


In order to know the coherence behavior of the Rydberg spin wave in the control channel $B$ during the operation of the photon switch, we have calculated Re$(\rho_{31}^B)$ as a function of the propagation distance $z$, with the results being presented in Fig.~\ref{delocalized}(c4) and Fig.~\ref{delocalized}(d4). We see that in the case of the facilitation regime Re$(\rho_{31}^B)$ becomes increasingly protected during the operation of the switch [Fig.~\ref{delocalized}(c4)]; however, in the case of the blockade regime  Re$(\rho_{31}^B)$ first increases due to resonance driving, and then rapidly decays to zero [Fig.~\ref{delocalized}(d4)] as $z$ increases. The reason for the  protected coherence during the switch operation in the facilitation regime is due to the fact that the control atom is driven away from the resonance with the choice of $\Delta_c=-V_{AB}$. Thereby, the coherence of the control atom is not diminished even in the case of the photon scattering.


\section{Summary and discussion}\label{conclusion}

We have investigated the FIT effect in a dual-channel Rydberg atom setting. It is found that optical properties of the probe photon in the target channel  can be actively controlled by varying laser detuning in the control channel. The FIT window emerges, and becomes wider and deeper as the inter-channel Rydberg interaction increases.  Using a dressed state approach, we have shown that the quantum interference pathways of the FIT are induced from the blockade and facilitated states. The FIT and its scaling on the interaction are examined both numerically and analytically. We have demonstrated that a contactless single-photon switch can be realized with FIT in the dual-channel system. Due to weak excitation under facilitated condition, coherence of the control channel is well protected during the operation of the switch.
This work moreover opens new opportunities to directly detect Rydberg facilitation using the optical transmission of a neighboring atom ensemble.

%

Beyond the single-photon situation considered here, multiple Rydberg excitations can form strongly correlated states in spatially large atomic gases (e.g. Rydberg excitation crystals~\cite{Pohl_Crystallization_2010} and quantum scarring states~\cite{turner2018weak}).  An interesting question is whether one could use this contactless multi-channel setting to probe Rydberg excitations and phases through Rydberg interaction induced optical responses~\cite{olmos_amplifying_2011,gunter_observing_2013}. This might provide new contactless detection methods without demolishing many-body coherence.

\begin{acknowledgements}
We thank useful discussions with Jianming Zhao, Lin Li, Yuechun Jiao, and Qi Zhang. ZB and GH acknowledge National Science Foundation of China (NSF) (11904104, 11975098), and the Shanghai Pujiang Program under grant No. 21PJ1402500. W. L. acknowledges support from the Engineering and Physical Sciences Research Council [grant no. EP/W015641/1]. The data that support this work are available in the Appendix of this article and at http://doi.org/10.17639/nott.7200.
\end{acknowledgements}

\appendix

\section{Multi-photon excitations in the control channel B}\label{ap1}
\setcounter{figure}{0}

One can also adopt an two- and multi-photon excitation scheme in the control channel $B$. Here we gives a simple description on a two-photon excitation scheme, shown in Fig.~\ref{SM_Model}.
\renewcommand\thefigure{A\arabic{figure}}
\begin{figure}[htb]
	\centering
	\includegraphics[width=0.5\textwidth]{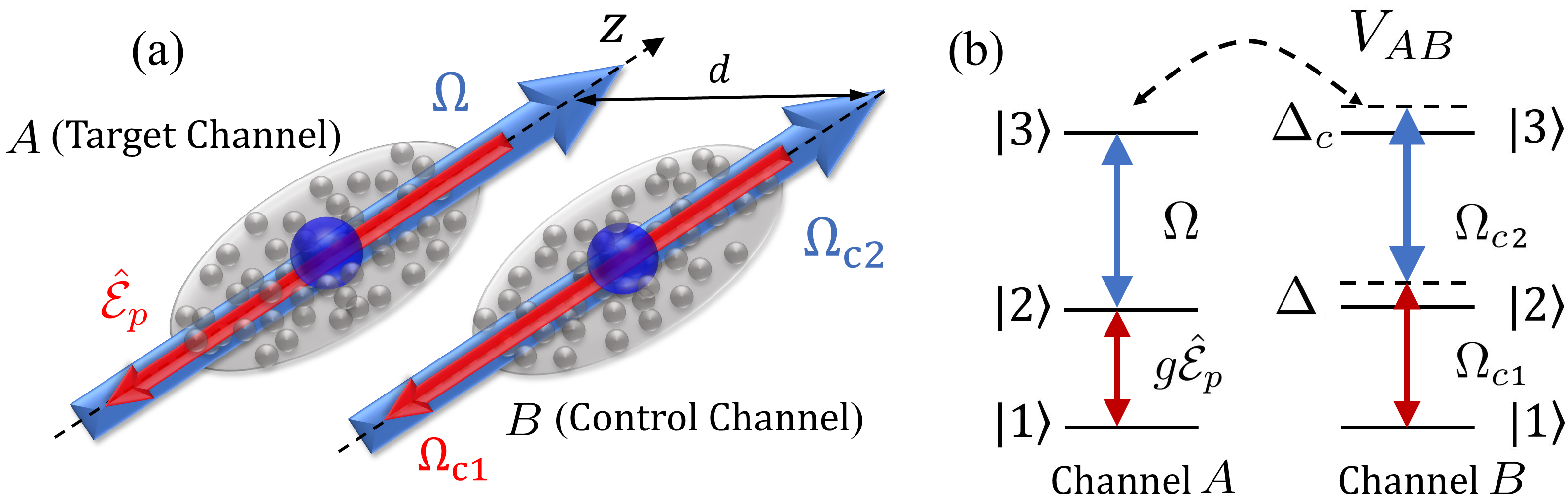}\\
	\caption{\footnotesize\textbf{Two-photon excitation scheme for the control channel $B$}. (a) The channel $A$ is the same as that of Fig.~\ref{Model}. The control channel $B$ is however driven by a two-photon process, where the transition $|1\rangle\leftrightarrow|2\rangle$\, ($|2\rangle\leftrightarrow|3\rangle$) is coupled by the control field of Rabi frequency $\Omega_{c1}$\, ($\Omega_{c2}$),
with detuning $\Delta$ ($\Delta_c$). (b) Level diagram of the system.} \label{SM_Model}
\end{figure}
In this scheme, the channel $A$ is the same as that in Fig.~\ref{Model} in the main text, but the control channel $B$ is a three-level system, where the first (second) control field with Rabi frequency $\Omega_{c1}$ ($\Omega_{c2}$) drives the transition $|1\rangle\leftrightarrow|2\rangle$ ($|2\rangle\leftrightarrow|3\rangle$)  with detuning $\Delta$ ($\Delta_c$). 
Under rotating wave approximation, the two-atom
Hamiltonian is given by $\hat{H}=\hat{H}_A+\hat{H}_B+\hat{H}_{\rm int}$, with
\begin{eqnarray*}\label{Hamiltonian}
\hat{H}_A&=&-\frac{\Omega_p}{2}\hat{\sigma}_{21}^A-\frac{\Omega}{2}\hat{\sigma}_{32}^A+{\rm H.c.}, \\
\hat{H}_B&=&-\left(\frac{\Omega_{c1}}{2}\hat{\sigma}_{21}^B+\frac{\Omega_{c2}}{2}\hat{\sigma}_{32}^B-\Delta\hat{\sigma}_{22}^B-\Delta_c\hat{\sigma}_{33}^B+{\rm H.c.}\right),\\
\hat{H}_{\rm int}&=&V_{AB}\hat{\sigma}_{33}^{A}\hat{\sigma}_{33}^{B}.
\end{eqnarray*}

\begin{figure}[htb]
	\centering \includegraphics[width=0.5\textwidth]{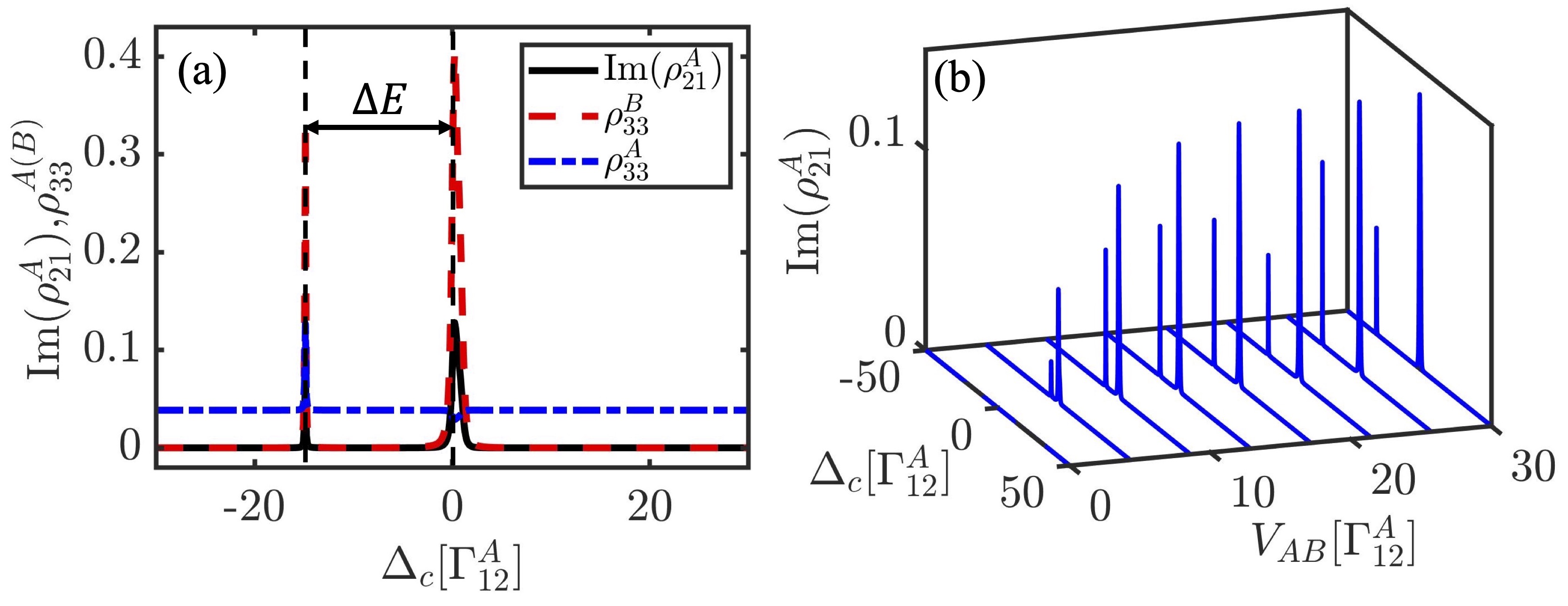}\\
	\caption{\footnotesize\textbf{Atomic coherence and Rydberg excitations in the dual-channel model when the control channel $B$ is driven by a two-photon process.} (a) FIT effect emerges in the absorption spectrum of the probe field, manifested by the imaginary part Im($\rho_{21}^A$) of the atomic coherence $\rho_{21}^A$ as a function of $\Delta_c$ (solid black line). Corresponding Rydberg excitations $\rho_{33}^{A}$ (dot-dashed blue line) and $\rho_{33}^{B}$ (dashed red line) are also plotted, where the channel $A$ is blockaded at $\Delta_c=0$, but facilitated around $\Delta_c=-V_{AB}$. (b) As $V_{AB}$ increases, the FIT transparency window becomes wider and deeper. Two peaks are visible even for weak interaction strength (e.g, $V_{AB}=5\Gamma_{12}^A$). System parameters used are $\Omega_{c1}=\Omega_p=0.8\Gamma_{12}^A$, $\Omega_{c2}=\Omega=4\Gamma_{12}^A$, and $\Delta=10\Gamma_{12}^A$.} \label{coherence_Multiphoton}
\end{figure}

The dynamical behavior of such a system can be obtained by numerically solving the corresponding master equation. Shown in Fig.~\ref{coherence_Multiphoton}(a)
is the result of Im($\rho_{21}^A$) (the imaginary part of the atomic coherence $\rho_{21}^A$) as a function of $\Delta_c$, plotted by the solid black line. The corresponding Rydberg excitations $\rho_{33}^{A}$ and $\rho_{33}^{B}$ are also illustrated by the dot-dashed blue line and dashed red line, respectively. We see that the channel $A$ is blockaded at $\Delta_c=0$, but facilitated near $\Delta_c=-V_{AB}$. Consequently, the FIT phenomenon can emerge in such a system. i.e. two absorption peaks emerge in the profile of Im($\rho_{21}^A$)
[Fig.~\ref{coherence_Multiphoton}(b)]. Moreover, the FIT transparency window becomes wider and deeper as $V_{AB}$ increases. Notice that, due to the large single-photon detuning $\Delta$ in the channel $B$ (which makes the effective Rabi frequency $\Omega_{\rm eff}=\Omega_{c1}\Omega_{c2}/(2\Delta)$ small), the linewidth of the absorption peak (proportional to $\Omega_{\rm eff}$) is much narrower than that obtained by using the one-photon excitation of the channel $B$ given in the main text. Thus by using such a scheme one can obtain a visible FIT doublet even for weak atomic interaction (e.g., $V_{AB}=5\Gamma_{12}^A$). This result is also shown in Fig.~\ref{Scaling}(c) (the blue triangle line) in the main text, and  agrees well with the dress-state theory described there.

\section{Atomic coherences in stationary states}\label{ap2}

The Heisenberg equations for operators $\hat{\sigma}_{13}^{A}$ and $\hat{\sigma}_{12}^{A}$  read
\begin{subequations}\label{Bloch_c}
\begin{eqnarray}
&& i\left(\frac{\partial }{\partial t}+\gamma_{12}^{A}\right)\hat{\sigma}_{12}^A
-\frac{\hat{\Omega}
_p}{2}(\hat{\sigma}_{22}^A-\hat{\sigma}_{11}^A)+\frac{\Omega^{\ast}}{2}
\hat{\sigma}_{13}^A=0,\label{Bloch_c1}\\
&& i\left(\frac{\partial }{\partial t}+\gamma_{13}^A\right)\hat{\sigma}_{13}^A-\frac{\hat{\Omega}
_p}{2}\hat{\sigma}_{23}^A+\frac{\Omega}{2}\hat{\sigma}_{12}^A+ V_{AB}\hat{\sigma}_{13}^{A}\hat{\sigma}_{33}^{B}=0. \label{Bloch_c2}\nonumber\\
\end{eqnarray}
\end{subequations}
Taking the average (trace) on the above equations (i.e. $\rho_{ab}^{A(B)}\equiv
\langle \sigma_{ba}^{A(B)}\rangle= {\rm Tr}[\rho \sigma_{ba}^{A(B)}]$)
and solving them for stationary states (i.e., $\partial/\partial t=0$), we obtain  the atomic coherence between
the states $|1\rangle$ and $|2\rangle$:
\begin{eqnarray}\label{Correlation_a1}
\rho_{21}^A= \frac{-2V_{AB}\,\rho_{31,33}^{AB}\Omega^\ast
+2i(\rho_{11}^A-\rho_{22}^A)\gamma_{13}^A\Omega_p+\rho_{32}^A\Omega^\ast\Omega_p}
{\Omega^2+4\gamma_{12}^{A}\gamma_{13}^A}.\nonumber\\
\end{eqnarray}
where $\rho_{31,33}^{AB}\equiv \langle\hat{\sigma}_{13}^{A}\hat{\sigma}_{33}^{B}\rangle$ is two-body DM element (or called two-body correlator). The terms related to the damping rate $\gamma_{13}^{A}$ are very small (e.g., 100${\rm \mu s}$) and hence can be neglected.
Moreover, since the probe field in the channel $A$ is weak, $\rho_{32}^A$ is negligible. Under these considerations, $\rho_{21}^A$ can be reduced to
\begin{eqnarray}\label{Correlation_a2}
\rho_{21}^A\simeq -\frac{2V_{AB}\,\rho_{31,33}^{AB}}{\Omega}.
\end{eqnarray}
We see that $\rho_{21}^A$  depends mainly on the two-body density matrix element  $\rho_{31,33}^{AB}$. From the result illustrated in Fig.~\ref{FIT_Windows}(d) of the main text, one can see that Eq.~(\ref{Correlation_a2}) can capture well with that obtained by solving the master equation numerically.

To get $\rho_{21}^A$, one must solve the equations of motion for two-body correlators $\rho_{ab,cd}^{AB}$ ($a,b,c,d=1,2,3$). For saving space, we do not list these equation here. For more detail, see \cite{Bai_Stable_2019}.

\section{Dressed-state picture for the FIT}\label{ap3}

As shown in Fig.~3(a) in the main text, the interaction-dressed subspace is considered with the basis $\{|2_A\rangle, |3_A\rangle, |1_B\rangle, |3_B\rangle\}$.  The Hamiltonian of this subspace reads
\begin{eqnarray*}\label{Hami0}
H_0=\begin{bmatrix}  0&-\Omega_c^{\ast}/2&-\Omega^{\ast}/2&0 \\
-\Omega_c/2& \Delta_{c} &0&-\Omega^{\ast}/2\\
-\Omega/2& 0 & 0&-\Omega_c^{\ast}/2\\
0&-\Omega/2&-\Omega_c/2&\Delta_{c}+V_{AB}
\end{bmatrix}\quad.
\end{eqnarray*}
After neglecting the small damping (i.e. spontaneous emission and dephasing) in the system, the eigenvalue (eigenstate) $\epsilon_j$\, $(|\psi_j\rangle)$ of the subspace can be obtained by diagonalizing the above Hamiltonian, with the result sketched in Fig.~\ref{Scaling}(b) of the main text.

In particular, for the zero eigenvalue ($\epsilon_j=0$), the probe field $\Omega_p$ can resonantly drive the transition from $|1_A\rangle$ to dressed subspace at the detuning defined by
\begin{eqnarray}
\Delta_{c}^{\pm}=-\frac{V_{AB}}{2}
\pm\frac{\sqrt{V_{AB}^2+[(\Omega^2-\Omega_{c}^2)/\Omega]^2}}{2}.
\end{eqnarray}
In the strong interaction regime (i.e. $V_{AB}\gg \Omega, \Omega_c$), we have $\Delta_{c}^{+}\simeq0$ and $\Delta_{c}^{-}\simeq-V_{AB}$. By adiabatically sweeping $\Delta_c$, the system can respectively excited to the Rydberg blockade and facilitation states, with the eigen state functions given by
\begin{subequations}
\begin{eqnarray}
\mid\Psi_{\rm E}\rangle&=&\frac{1}{\sqrt{2}}(\mid 3_A1_B\rangle-\mid 2_A3_B\rangle),\,\,\,\,{\rm for}\,\, \Delta_{c}=\Delta_c^{+},\\
\mid\Psi_{\rm F}\rangle&=&\frac{1}{\sqrt{2}}(\mid 3_A3_B\rangle-\mid 2_A1_B\rangle),\,\,\,\,{\rm for}\,\, \Delta_{c}=\Delta_c^{-}.
\end{eqnarray}
\end{subequations}
However, due to the large decay of the level $|2_A\rangle$ (which is not included in the calculation on the eigenstates), entangled state $|\Psi_E\rangle$ rapidly evolves to trivial blockade state $|\Psi_B\rangle=|1_A3_B\rangle$. The energy gap between $|\Psi_E\rangle$ and  $|\Psi_B\rangle$ is
\begin{eqnarray}\label{DeltaE}
\Delta E =\sqrt{V_{AB}^2+(\Omega-\Omega_c^2/\Omega)^2}\simeq V_{AB},
\end{eqnarray}
which is plotted (with dashed red line) in Fig.~\ref{Scaling}(c) of the main text. We find that the result of this dressed-state approach can capture the numerical data very well in the interaction dominant regime.

\begin{figure}
\centering
\includegraphics[width=0.5\textwidth]{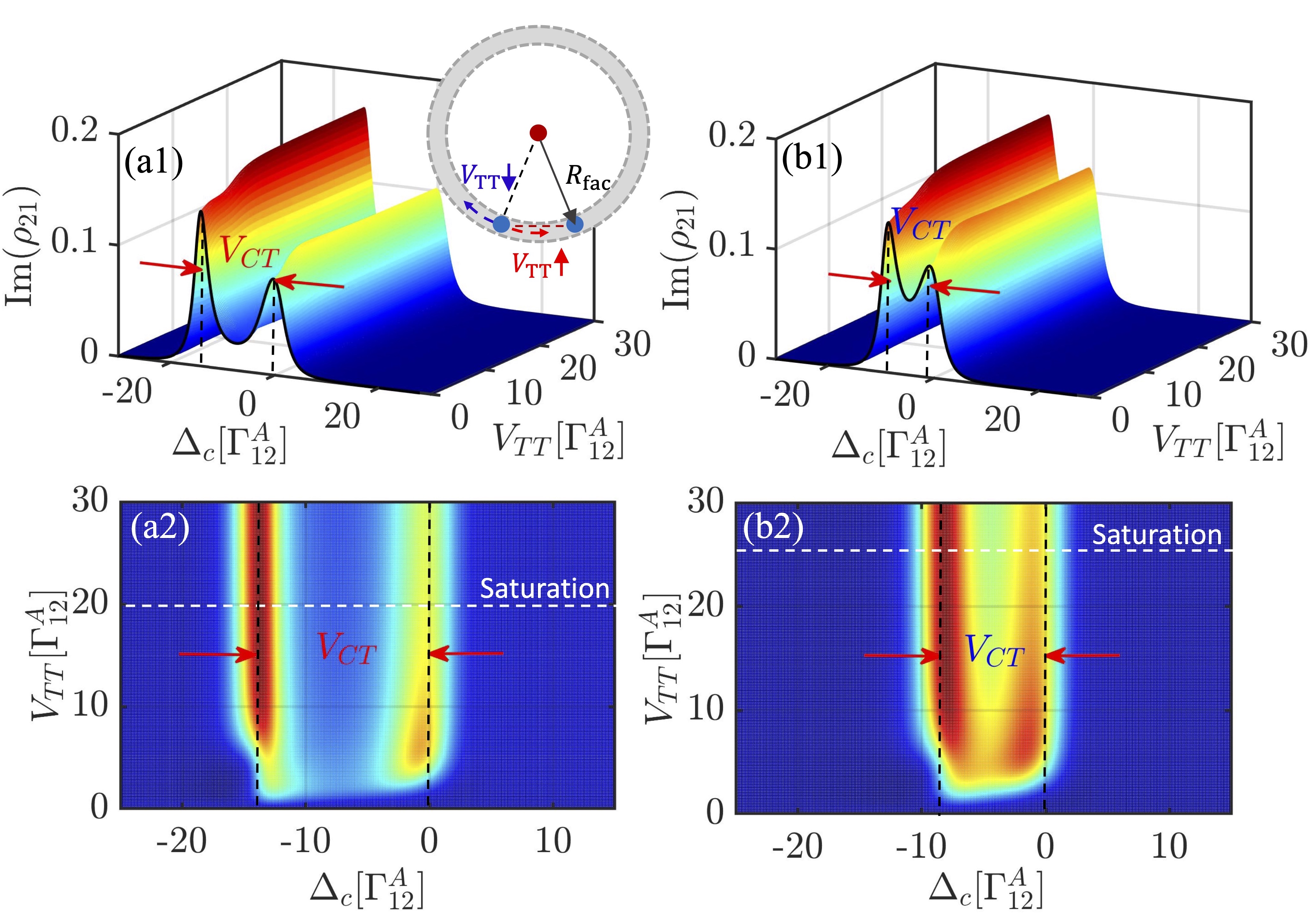}\\
\caption{\footnotesize \textbf{Three-atom model}.
(a1) FIT spectrum Im($\rho_{21}$) for the target channel as functions of $\Delta_c$ and $V_{TT}$, which exhibits double peaks even for a strong $V_{TT}$, with $V_{CT}=15\Gamma_{12}^A$. The inset shows the geometry of the system, in which the control channel (red circle) locates at the center of a ring (with radius $R_{\rm fac}$), and the target channels (two blue circles) locate on the shell.
(a2)  The corresponding contour map as functions of $\Delta_c$ and $V_{TT}$ is sketched
with Im$[\rho_{21}-\rho_{12}(V_{TT}=0)]$. It is clear to see that the results reach saturation as $V_{TT}$ increases.
(b1) and (b2)  are similar to (a1) and (a2) but with $V_{CT}=10\Gamma_{12}^A$.}
\label{3AM}
\end{figure}


\section{FIT in the three-atom model}\label{ap4}


To explore the influence of the target-target (TT) interaction (described by $V_{TT}$) in systems with multi-channels, here we consider a simple three-atom scenario with one control channel and two target channels; see the inset of Fig.~\ref{3AM}(a1) and the corresponding caption. For simplicity, we assume that the control-target (CT) interaction $V_{CT}$ is fixed, but $V_{TT}$ can be changed by varying position of target channels on the shell.

The FIT behavior in such a system is investigated by solving the corresponding Maxwell-Bloch equations of the model, with results given by in Fig.~\ref{3AM}.
Panel (a1) shows the FIT spectrum Im($\rho_{21}$) for the target-channel  as a function of $\Delta_c$ and $V_{TT}$, with $V_{CT}=15\Gamma_{12}^A$. We see that it exhibits two peaks even $V_{TT}$ is strong. Illustrated in panel (a2) is the contour map of $V_{TT}$ as a function of $\Delta_c$, sketched with Im$[\rho_{21}-\rho_{12}(V_{TT}=0)]$. One sees that it reaches a saturation as $V_{TT}$ increases. For comparison, similar result for case of $V_{CT}=10\Gamma_{12}^A$ is also provided; one see that FIT can also saturate but need larger $V_{TT}$ [see panels (b1) and (b2)]. From these results, we conclude that the TT interaction has a minor influence on the occurrence of the FIT. In fact, under the condition of $\Delta\simeq-V_{CT}$, the control atom can facilitate the all atoms located on the shell of the blockade sphere, and it can lead to a self-organized criticality, which has been observed in a recent study~\cite{Signatures_Helmrich_2020}.
This example shows that the number of target channels  is largely not limited in the FIT model as long as the conditions are met. In practice, of course, the number of target channels will be limited by physical dimensions of the channel. Hence future experiments can explore such practical limitations.
%
%




%

\end{document}